%% file: arxiv_version.tex
\documentclass[10pt, a4paper, twocolumn]{extarticle}
\pdfoutput=1

\newcommand{\movieref}[1]{ and Video S#1}
\usepackage{titlefoot}
\usepackage[activate={true,nocompatibility},final,tracking=true,kerning=true,spacing=true,factor=1100,stretch=10,shrink=10]{microtype}
\usepackage{amsmath}
\usepackage{amssymb}
\usepackage{hyperref}
\hypersetup{
    colorlinks,
    linkcolor={blue!80!black},
    citecolor={blue!80!black},
    urlcolor={blue!80!black}
}

\usepackage[capitalise]{cleveref}
\usepackage{here}
\usepackage{geometry}
\usepackage[numbers, compress]{natbib}
\usepackage{siunitx}
\usepackage{xspace}
\usepackage{todonotes}
\include{tex_files/newcommands}

\usepackage{graphicx}
\usepackage{braket}
\usepackage{multirow}                 
\usepackage{booktabs}

\usepackage{caption}
\usepackage{setspace}
\title{\Huge Stochasticity from function -- \\why the Bayesian brain may need no noise}

\author{
    Dominik Dold\textsuperscript{a,b,1,2}, Ilja Bytschok\textsuperscript{a,1}, Akos F. Kungl\textsuperscript{a,b}, Andreas Baumbach\textsuperscript{a},\\
    Oliver Breitwieser\textsuperscript{a}, Walter Senn\textsuperscript{b}, Johannes Schemmel\textsuperscript{a}, Karlheinz Meier\textsuperscript{a}, Mihai A. Petrovici\textsuperscript{a,b,1,2}
    \vspace{10pt}\\
    \textsuperscript{a}\textit{Kirchhoff-Institute for Physics, Heidelberg University.}\\
    \textsuperscript{b}\textit{Department of Physiology, University of Bern.}\\
    \textsuperscript{1}\textit{Authors with equal contributions.}\\
    \textsuperscript{2}\textit{Corresponding authors}: dodo@kip.uni-heidelberg.de, petrovici@pyl.unibe.ch. 
    }

\date{\today}

\begin{document}
\onecolumn

\maketitle
\vspace{1.25cm}
\section*{Abstract}
\input{tex_files/abstract}
\clearpage
\twocolumn
\section{Introduction}
\input{tex_files/introduction}
%
\input{tex_files/methods}
\section{Results}
\input{tex_files/noise_statistics}
\input{tex_files/sampling_without_noise}
\input{tex_files/neuromorphic}
\input{tex_files/applications}
\input{tex_files/discussion}
\input{tex_files/conclusion}
\input{supplement/calc_details}
\section{Acknowledgments}
\input{tex_files/acknowledgments}
\section{References}
\renewcommand{\bibsection}{}
\bibliography{tex_files/mybib}
\renewcommand{\thefigure}{S\arabic{figure}}
\setcounter{figure}{0}
\section{Supporting information}
\input{supplement/sim_details}
\subsection{Neuron parameters}
\input{supplement/Table_S1}
\input{supplement/Table_S2}
\input{supplement/movies}
\clearpage
\onecolumn
\subsection{Supporting figures}
\input{supplement/SI_figures}

\end{document}

%% file: tex_files/newcommands.tex
\renewcommand{\and}{{\mathrm{and}}\xspace}

\newcommand{\bs}[1]{\boldsymbol{#1}}

\newcommand{\DKL}[2]{D_\mathrm{KL}\left(#1\parallel#2\right)}

\newcommand{\CC}{\rho\xspace}

\newcommand{\Cm}{{C_\mathrm{m}}\xspace}
\newcommand{\col}{{\mathrm{col}}\xspace}

\newcommand{\corr}{{\mathcal C}\xspace}

\newcommand{\DKLsolo}{{D_\mathrm{KL}}\xspace}
\newcommand{\El}{{E_\mathrm{l}}\xspace}

\newcommand{\Erev}{E^\mathrm{rev}\xspace}
\newcommand{\Ereve}{{E^\mathrm{rev}_\mathrm{e}}\xspace}
\newcommand{\Erevi}{{E^\mathrm{rev}_\mathrm{i}}\xspace}

\newcommand{\gl}{{g_\mathrm{l}}\xspace}

\newcommand{\gsyn}{g^\mathrm{syn}\xspace}

\newcommand{\gtot}{{g^\mathrm{tot}}\xspace}

\newcommand{\pnet}{p^\mathrm{net}}

\newcommand{\ptarget}{p^\mathrm{target}}

\newcommand{\taueff}{{\tau_\mathrm{eff}}\xspace}

\newcommand{\taurec}{{\tau_\mathrm{rec}}\xspace}
\newcommand{\tauref}{{\tau_\mathrm{ref}}\xspace}

\newcommand{\tausyn}{\tau^\mathrm{syn}\xspace}

\newcommand{\ueff}{{u_\mathrm{eff}}\xspace}

\newcommand{\USE}{{U_\mathrm{SE}}\xspace}

\tikzstyle{neuron}=[circle,fill=black!25,minimum size=21pt,inner sep=0pt]
\tikzstyle{visible neuron}=[neuron, fill=green!50]
\tikzstyle{hidden neuron}=[neuron, fill=orange!75]

%% file: tex_files/abstract.tex
An increasing body of evidence suggests that the trial-to-trial variability of spiking activity in the brain is not mere noise, but rather the reflection of a sampling-based encoding scheme for probabilistic computing.
Since the precise statistical properties of neural activity are important in this context, many models assume an ad-hoc source of well-behaved, explicit noise, either on the input or on the output side of single neuron dynamics, most often assuming an independent Poisson process in either case.
However, these assumptions are somewhat problematic: neighboring neurons tend to share receptive fields, rendering both their input and their output correlated; at the same time, neurons are known to behave largely deterministically, as a function of their membrane potential and conductance.
We suggest that spiking neural networks may have no need for noise to perform sampling-based Bayesian inference.
We study analytically the effect of auto- and cross-correlations in functional Bayesian spiking networks and demonstrate how their effect translates to synaptic interaction strengths, rendering them controllable through synaptic plasticity.
This allows even small ensembles of interconnected deterministic spiking networks to simultaneously and co-dependently shape their output activity through learning, enabling them to perform complex Bayesian computation without any need for noise, which we demonstrate in silico, both in classical simulation and in neuromorphic emulation.
These results close a gap between the abstract models and the biology of functionally Bayesian spiking networks, effectively reducing the architectural constraints imposed on physical neural substrates required to perform probabilistic computing, be they biological or artificial.

%% file: tex_files/introduction.tex
An ubiquitous feature of in-vivo neural responses is their stochastic nature \cite{henry1973orientation, schiller1976short, vogels1989response, snowden1992response, arieli1996dynamics, azouz1999cellular}.
The clear presence of this variability has spawned many functional interpretations, with the Bayesian-brain hypothesis arguably being the most notable example \cite{rao2002probabilistic, kording2004bayesian, brascamp2006time, deco2009stochastic, fiser2010statistically, maass2016searching}.
Under this assumption, the activity of a neural network is interpreted as representing an underlying (prior) probability distribution, with sensory data providing the evidence needed to constrain this distribution to a (posterior) shape that most accurately represents the possible states of the environment given the limited available knowledge about it.

Neural network models have evolved to reproduce this kind of neuronal response variability by introducing noise-generating mechanisms, be they extrinsic, such as Poisson input \cite{stein1967some,brunel2000dynamics, fourcaud2002dynamics, gerstner2014neuronal} or fluctuating currents \cite{smetters1996synaptic, maass1998dynamic, steinmetz2000subthreshold, yarom2011voltage, moreno2014poisson,neftci2016stochastic}, or intrinsic, such as stochastic firing \cite{stevens1996integrate, plesser2000noise, chichilnisky2001simple, gerstner2002spiking, dayan2003theoretical, ostojic2011spiking} or membrane fluctuations \cite{orban2016neural,aitchison2016hamiltonian}. 

However, while representing, to some degree, reasonable approximations, none of the commonly used sources of stochasticity is fully compatible with biological constraints.
Contrary to the independent white noise assumption, neuronal inputs are both auto- and cross-correlated to a significant degree \cite{deger2012statistical, nelson1992spatial, averbeck2006neural, salinas2001correlated, segev2002long, fiser2004small, rosenbaum2014correlated}, with obvious consequences for a network's output statistics \cite{moreno2008theory}.
At the same time, the assumption of intrinsic neuronal stochasticity is at odds with experimental evidence of neurons being largely deterministic units \cite{mainen1995reliability, zador1998impact, rauch2003neocortical}.
Although synaptic transmissions from individual release sites are stochastic, averaged across multiple sites, contacts and connections, they largely average out \cite{markram2015reconstruction}.
Therefore, it remains an interesting question how cortical networks that use stochastic activity as a means to perform probabilistic inference can realistically attain such apparent randomness in the first place.

We address this question within the normative framework of sampling-based Bayesian computation \cite{orban2016neural, buesing2011neural, petrovici2016stochastic, pecevski2011probabilistic, probst2015probabilistic, leng2018spiking}, in which the spiking activity of neurons is interpreted as Markov Chain Monte Carlo sampling from an underlying distribution over a high-dimensional binary state space.
In contrast to other work on deterministic chaos in functional spiking networks, done mostly in the context of reservoir computing (e.g., \cite{monteforte2012dynamic, pyle2017spatiotemporal}), we provide a stringent connection to the spike-based representation and computation of probabilities, as well as the synaptic plasticity required for learning the above.
We demonstrate how an ensemble of dynamically fully deterministic, but functionally probabilistic networks, can learn a connectivity pattern that enables probabilistic computation with a degree of precision that matches the one attainable with idealized, perfectly stochastic components.
The key element of this construction is self-consistency, in that all input activity seen by a neuron is the result of output activity of other neurons that fulfill a functional role in their respective subnetworks.
The present work supports probabilistic computation in light of experimental evidence from biology and suggests a resource-efficient implementation of stochastic computing by completely removing the need for any form of explicit noise.

%% file: tex_files/methods.tex
\section{Methods}

    \subsection{Neuron model and simulation details}
    
        We consider deterministic Leaky Integrate-and-Fire (LIF) neurons with conductance-based synapses and dynamics described by 
        \begin{align}
        \Cm \frac{\mathrm{d}u_k}{\mathrm{d}t} &=\gl \left(\El - u_k\right) + \sum_{x\in \{\mathrm{e},\mathrm{i}\}} \gsyn_{k,x} (\Erev_x - u_k) \, , \label{eq:LIF} \\
        \gsyn_{k,x}(t) &= \!\!\!\!\!\! \sum_{\mathrm{synapses\hspace{.5mm} j}} \hspace{.5mm} \sum_{\mathrm{spikes\hspace{.75mm} s}} \! w_{kj}\ \theta(t-t_s)\ \exp{(-\frac{t-t_s}{\tausyn})} \, , \\
        u_k(t_s) &\geq \vartheta \hspace{2mm} \Rightarrow \hspace{2mm} u_k(t \in (t_s, t_s+\tauref]) = \varrho \, ,
        \end{align}
        with membrane capacitance $\Cm$, leak conductance $\gl$, leak potential $\El$, excitatory and inhibitory reversal potentials $\Erev_{\mathrm{e}/\mathrm{i}}$ and conductances $\gsyn_{k,\mathrm{e}/\mathrm{i}}$, synaptic strength $w_{kj}$, synaptic time constant $\tausyn$ and $\theta(t)$ the Heaviside step function.
        For $\gsyn_{k,\mathrm{e}/\mathrm{i}}$, the first sum covers all synaptic connections projecting to neuron $k$.
        A neuron spikes at time $t_s$ when its membrane potential crosses the threshold $\vartheta$, after which it becomes refractory.
        During the refractory period $\tauref$, the membrane potential is clamped to the reset potential $\varrho$.
        We have chosen the above model because it provides a computationally tractable abstraction of neurosynaptic dynamics \cite{rauch2003neocortical}, but our general conclusions are not restricted to these specific dynamics.

        We further use the short-term plasticity mechanism described in \cite{fuhrmann2002coding} to modulate synaptic interaction strengths with an adaptive factor $U_\mathrm{SE}\times R(t)$, where the time-dependence is given by\footnote{In \cite{fuhrmann2002coding} the postsynaptic response only scales with $R(t)$, whereas here we scale it with $U_\mathrm{SE} \times R(t)$.}
        \begin{equation}
        \frac{\mathrm{d}R}{\mathrm{d}t} = \frac{1-R}{\taurec} - \USE R\delta(t-t_s)\,,\hspace{4mm} \USE\,,\,R \in [0,1]\,,
        \end{equation}
        where $\delta(t)$ is the Dirac delta function, $t_s$ denotes the time of a presynaptic spike, which depletes the reservoir $R$ by a fraction $U_\mathrm{SE}$, and $\tau_\mathrm{rec}$ is the time scale on which the reservoir $R$ recovers.
        This enables a better control over the inter-neuron interaction, as well as over the mixing properties of our networks \cite{leng2018spiking}.
        
        Background input, such as spikes from a Poisson source, enters \cref{eq:LIF} as synaptic input, but without short-term plasticity (as in \cite{petrovici2016stochastic}) to facilitate the mathematical analysis (see Supporting information for more details).
        
        All simulations were performed with the network specification language PyNN 0.8 \cite{pynn2008} and the spiking neural network simulator NEST 2.4.2 \cite{Gewaltig:NEST}.

    \subsection{Sampling framework}
    
\begin{figure}
        \centering
        \includegraphics[width=1.\linewidth]{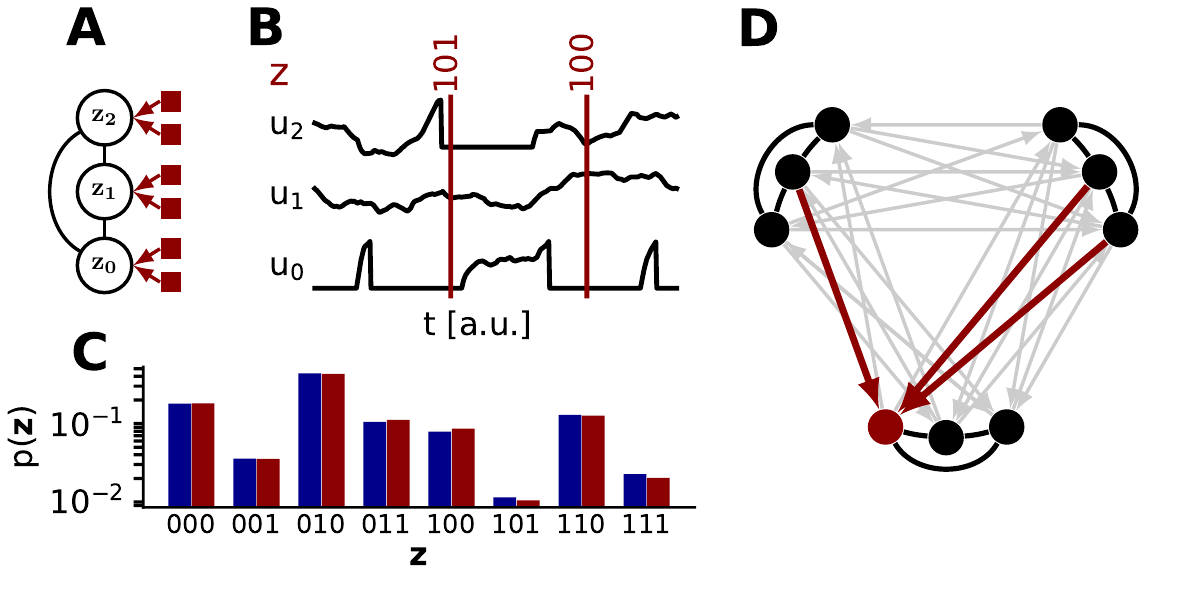}
        \caption{
            Sampling spiking networks (SSNs) with and without explicit noise.
            \textbf{(A)} Schematic of a sampling spiking network, where each neuron (circles) encodes a binary random variable $z_i \in \{0,1\}$.
                         In the original model, neurons were rendered effectively stochastic by adding external Poisson sources of high-frequency balanced noise (red boxes).
            \textbf{(B)} A neuron represents the state $z_k=1$ when refractory and $z_k=0$ otherwise.
            \textbf{(C)} The dynamics of neurons in an SSN can be described as sampling (red bars) from a target distribution (blue bars).
            \textbf{(D)} Instead of using Poisson processes as a source of explicit noise, we replace the Poisson input with spikes coming from other networks performing spike-based probabilistic inference by creating a sparse, asymmetric connectivity matrix between several SSNs.
                         For instance, the red neuron receives not only information-carrying spikes from its home network (black lines), but also spikes from the other two SSNs as background (red arrows), and in turn projects back towards these networks.
                         Other such background connections are indicated in light gray. 
        }
        \label{fig:fig1}
        \end{figure}

        As a model of probabilistic inference in networks of spiking neurons, we adopt the framework introduced in \cite{petrovici2016stochastic, probst2015probabilistic}.
        There, the neuronal output becomes stochastic due to a high-frequency bombardment of excitatory and inhibitory Poisson stimuli (\cref{fig:fig1}A), elevating neurons into a high-conductance state (HCS) \cite{destexhe2003high,kumar2008high}, where they attain a high reaction speed due to a reduced effective membrane time constant.
        Under these conditions, a neuron's response (or activation) function becomes approximately logistic and can be represented as $\varphi(\mu) = \left(1+\exp{(-(\mu-u_0)/\alpha})\right)^{-1}$ with inverse slope $\alpha$ and inflection point $u_0$.
        Together with the mean free membrane potential $\mu$ and the mean effective membrane time constant $\taueff$ (see \cref{eq:SI_mu,eq:SI_taueff}), the scaling parameters $\alpha$ and $u_0$ are used to translate the weight matrix $\bs W$ and bias vector $\bs b$ of a target Boltzmann distribution $\ptarget_{\bs z} = p({\bs z}) \propto \exp\left( \frac{1}{2} \bs z^T \bs W \bs z + \bs z^T \bs b \right)$ with binary random variables $\bs z \in \{0, 1\}^n$ to synaptic weights and leak potentials in a sampling spiking network (SSN):
        \begin{align}
            w_{kj} &= \frac{\alpha W_{kj} \Cm \frac{\tauref}{\tausyn} \left(1-\frac{\tausyn}{\taueff}\right) \left(\Erev_{kj} - \mu\right)^{-1}}{ \left[\tausyn \left(e^{-\frac{\tauref}{\tausyn}} - 1 \right) - \taueff \left( e^{- \frac{\tauref}{\taueff}} - 1 \right) \right]} \label{eq:wtransl} \, , \\
            \bs \El &= \frac{\tau_\mathrm{m}}{\taueff} (\alpha \bs b + u_0) - \sum_{x\in \{\mathrm{e},\mathrm{i}\}} \frac{\braket{\gsyn_x}}{\gl} \Erev_x \label{eq:btransl}\,,
        \end{align}
        where $w_{kj}$ is the synaptic weight from neuron $j$ to neuron $k$, $\bs\El$ a vector containing the leak potentials of all neurons, $\bs b$ the corresponding bias vector, $\Erev_{kj} \in \{\Ereve, \Erevi$\}, depending on the nature of the respective synapse, and $\tau_\mathrm{m} = \frac{\Cm}{\gl}$ (see \cref{eq:SI_motivate_transl} to \cref{eq:SI_w_rule} for a derivation). This translation effectively enables sampling from $\ptarget_{\bs z}$, where a refractory neuron is considered to represent the state $z_k=1$ (see \cref{fig:fig1}B,C).
               
    \subsection{Measures of network performance}
    
        To assess how well a sampling spiking network (SSN) samples from its target distribution, we use the Kullback-Leibler divergence \cite{kullback1951information}
        \begin{equation}
        \DKL{\pnet}{\ptarget} = \sum_{\bs z} \pnet_{\bs z} \ln{\left(\frac{\pnet_{\bs z}}{\ptarget_{\bs z}}\right)}\,,
        \end{equation}
        which is a measure for the similarity between the sampled distribution $\pnet$ and the target distribution $\ptarget$.
        For inference tasks, we determine the network's classification rate on a subset of the used data set which was put aside during training.
        Furthermore, generative properties of SSNs are investigated either by letting the network complete partially occluded examples from the data set or by letting it generate new examples.

    \subsection{Learning algorithm}
    
        Networks were trained with a Hebbian wake-sleep algorithm
        \begin{align}
        \Delta {\bs W}_{ij} &= \eta \left[ \ptarget_{z_i = 1, z_j = 1} - \pnet_{z_i = 1, z_j = 1} \right] \label{eq:CDw}\,, \\
        \Delta {\bs b}_i &= \eta \left[ \ptarget_{z_i = 1} - \pnet_{z_i = 1} \right] \label{eq:CDb} \,,
        \end{align}
        which minimizes the $\DKL{\pnet}{\ptarget}$ \cite{ackley1985learning}.
        $\eta$ is a learning rate (see Supporting information for used hyperparameters).
        For high-dimensional datasets (e.g. handwritten letters and digits), Boltzmann machines were trained with the CAST algorithm \cite{salakhutdinov2010learning}, a variant of wake-sleep with a tempering scheme, and then translated to SSN parameters with \cref{eq:wtransl,eq:btransl} instead of training the SSNs directly to reduce simulation time.

    \subsection{Experiments and calculations} 

        Details to all experiments as well as additional figures and captions to videos can be found in the Supporting information. Detailed calculations are presented at the end of the main text.

%% file: tex_files/noise_statistics.tex
We approach the problem of externally-induced stochasticity incrementally.
Throughout the remainder of the manuscript, we discern between background input, which is provided by other functional networks, and explicit noise, for which we use the conventional assumption of Poisson spike trains.
We start by analyzing the effect of correlated background on the performance of SSNs.
We then demonstrate how the effects of both auto- and cross-correlated background can be mitigated by Hebbian plasticity.
This ultimately enables us to train a fully deterministic network of networks to perform different inference tasks without requiring any form of explicit noise.
This is first shown for larger ensembles of small networks, each of which receives its own target distribution, which allows a straightforward quantitative assessment of their sampling performance $\DKL{\pnet}{\ptarget}$.
We study the behavior of such ensembles both in computer simulations and on mixed-signal neuromorphic hardware.
Finally, we demonstrate the capability of our approach for truly functional, larger-scale networks, trained on higher-dimensional visual data.

\subsection{Background autocorrelations}

    \input{tex_files/fig2.tex}

    Unlike ideal Poisson sources, single spiking neurons produce autocorrelated spike trains, with the shape of the autocorrelation function (ACF) depending on their refractory time $\tauref$ and mean spike frequency $\bar r = p(z=1) \tauref^{-1}$.
    For higher output rates, spike trains become increasingly dominated by bursts, i.e., sequences of equidistant spikes with an interspike interval (ISI) of $\mathrm{ISI} \approx \tauref$.
    These fixed structures also remain in a population, since the population autocorrelation is equal to the averaged ACFs of the individual spike trains.
    
    We investigated the effect of such autocorrelations on the output statistics of SSNs by replacing the Poisson input in the ideal model with spikes coming from other SSNs.
    As opposed to Poisson noise, the autocorrelation $\corr (S_x, S_x, \Delta) = \frac{\braket{S_x(t)S_x(t+\Delta)}-\braket{S_x}^2}{\mathrm{Var}(S_x)}$ of the SSN-generated (excitatory or inhibitory) background $S_x,\,x \in \{\mathrm{e},\mathrm{i}\}$ (\cref{fig:fig2}B) is non-singular and influences the free membrane potential (FMP) distribution (\cref{fig:fig2}C) and thereby activation function (\cref{fig:fig2}D) of individual sampling neurons.
    With increasing firing rates (controlled by the bias of the neurons in the background SSNs), the number of significant peaks in the ACF increases as well (see \cref{eq:SI_autocorr}):
    \begin{equation}
        \corr(S_x, S_x, n\tauref) \approx \sum_{k=1}^\infty e^{k \ln{\bar{p}}} \delta\big([n-k]\tauref\big)\,,
        \label{eqn:acf}
    \end{equation}
    where $\bar{p}$ is the probability for a burst to start.
    This regularity in the background input manifests itself in a reduced width $\sigma'$ of the FMP distribution (see \cref{eq:SI_width_rescale})
    \begin{equation}
        f(u_i^\mathrm{free}) \sim \mathcal N (\mu' = \mu, \sigma' = \sqrt{\beta}\sigma ) \label{eqn:fmp}
    \end{equation}
    with a scaling factor $\sqrt{\beta}$ that depends on the ACF, which in turn translates to a steeper activation function (see \cref{eq:SI_inverse_slope,eq:SI_rescale_inv_slope})
    \begin{equation}
        p(z_i=1) \approx \int_\vartheta^\infty f(u) \mathrm du \approx \varphi(\mu) \bigg|_{u_0' = u_0, \alpha' = \sqrt{\beta} \alpha} \, ,
        \label{eqn:actfct-approx}
    \end{equation}
    with inflection point $u_0'$ and inverse slope $\alpha'$. Thus, autocorrelations in the background input lead to a reduced width of the FMP distribution and hence to a steeper activation function compared to the one obtained using uncorrelated Poisson input.
    For a better intuition, we used an approximation of the activation function of LIF neurons, but the argument also holds for the exact expression derived in \cite{petrovici2016stochastic}, as verified by simulations (\cref{fig:fig2}D).
    
    Apart from the above effect, the background autocorrelations do not affect neuron properties that depend linearly on the synaptic noise input, such as the mean FMP and the inflection point of the activation function (equivalent to zero bias).
    Therefore, the effect of the background autocorrelations can be functionally reversed by rescaling the functional (from other neurons in the principal SSN) afferent synaptic weights by a factor equal to the ratio between the new and the original slope $\alpha'/\alpha$ (\cref{eq:wtransl,eq:btransl}), as shown in \cref{fig:fig2}E.

\subsection{Background cross-correlations}
    
    In addition to being autocorrelated, background input to pairs of neurons can be cross-correlated as well, due to either shared inputs or synaptic connections between the neurons that generate said background.
    These background cross-correlations can manifest themselves in a modified cross-correlation between the outputs of neurons, thereby distorting the distribution sampled by an SSN.
    
    However, depending on the number and nature of presynaptic background sources, background cross-correlations may cancel out to a significant degree.
    The correlation coefficient (CC) of the FMPs of two neurons fed by correlated noise amounts to (see \cref{eq:SI_crosscorr})
    \begin{align}
        \CC (u_i^\mathrm{free}, u_j^\mathrm{free}) \propto \sum_{l,m}& w_{il} w_{jm} \big(E_{il}^\mathrm{rev}- \mu_i\big)\big(E_{jm}^\mathrm{rev}- \mu_j\big)\label{eqn:cc} \\ 
        \cdot \ \int \mathrm{d}\Delta \ \lambda_{li, mj}\ \corr & \left(S_{l\mathrm{,i}}, S_{m\mathrm{,j}}, \Delta\right) \widetilde{\corr}\left(\kappa, \kappa, \Delta\right)\,, \nonumber
    \end{align}
    where $l$ sums over all background spike trains $S_{l,i}$ projecting to neuron $i$ and $m$ sums over all background spike trains $S_{m,j}$ projecting to neuron $j$. $\widetilde{\corr}\left(\kappa, \kappa, \Delta\right)$ is the unnormalized autocorrelation function of the postsynaptic potential (PSP) kernel $\kappa$, i.e., $\widetilde{\corr}\left(\kappa, \kappa, \Delta\right) = \braket{\kappa(t)\kappa(t+\Delta)}$, and $\corr\left(S_{l\mathrm{,i}}, S_{m\mathrm{,j}}, \Delta\right)$ the cross-correlation function of the background inputs. 
    $\lambda_{li, mj}$ is given by $\lambda_{li, mj} = \sqrt{\mathrm{Var}\left(S_{l\mathrm{,i}}\right)\mathrm{Var}\left(S_{m\mathrm{,j}}\right)}$.
    The background cross-correlation is gated into the cross-correlation of FMPs by the nature of the respective synaptic connections: if the two neurons connect to the cross-correlated inputs by synapses of different type (one excitatory, one inhibitory), the sign of the CC is switched (\cref{fig:fig2}F).
    However, individual contributions to the FMP CC also depend on the difference of the mean free membrane potential and the reversal potentials, so the gating of cross-correlations is not symmetric for excitatory and inhibitory synapses.
    Nevertheless, if the connectivity statistics (in-degree and synaptic weights) from the background sources to an SSN are chosen appropriately and enough presynaptic partners are available, the total pairwise cross-correlation between neurons in an SSN cancels out on average, leaving the sampling performance unimpaired (\cref{fig:fig2}G).
    Note that this way of reducing cross-correlations is independent of the underlying weight distribution of the networks providing the background; the required cross-wiring of functional networks could therefore, in principle, be encoded genetically and does not need to be learned.
    Furthermore, a very simple cross-wiring rule, i.e., independently and randomly determined connections, already suffices to accomplish low background cross-correlations and therefore reach a good sampling performance. 
    
    Whereas this method is guaranteed to work in an artificial setting, further analysis is needed to assess its compatibility with the cortical connectome with respect to connectivity statistics or synaptic weight distributions.
    However, even if cortical architecture prevents a clean implementation of this decorrelation mechanism, SSNs can themselves compensate for residual background cross-correlations by modifying their parameters, similar to the autocorrelation compensation discussed above.
    
    To demonstrate this ability, we need to switch from the natural state space of neurons $\bs z \in \{0, 1\}^N$ to the more symmetric space $\bs z' \in \{-1, 1\}^N$.\footnote{The $z=0$ state for a silent neuron is arguably more natural, because it has no effect on its postsynaptic partners during this state. In contrast, $z \in \{-1, 1\}$ would, for example, imply efferent excitation upon spiking and constant efferent inhibition otherwise.}
    By requiring $p(\bs z') \overset{!}{=} p(\bs z)$ to conserve state probabilities (and thereby also correlations), the desired change of state variables $\bs z' = 2 \bs z - \bs 1$ can be achieved with a linear parameter transformation (see \cref{SI:eq1,SI:eq2}):
    \begin{equation}
        \bs W' = \frac 1 4 \bs W \; \mathrm{and} \; \bs b' = \frac 1 2 \bs b + \frac 1 4 \sum_i \col_i \bs W \, .
    \end{equation}
    In the $\{-1, 1\}^N$ state space, both synaptic connections $w'_{ij}$ and background cross-correlations $\CC(S_i, S_j)$ shift probability mass between mixed states $(z_i, z_j) = \pm(1,-1)$ and aligned states $(z_i, z_j) = \pm(1,1)$ (see Supporting information, \cref{fig:figS1}).
    Therefore, by adjusting $\bs b$ and $\bs W$, it is possible to find a $\bs W'$ (\cref{fig:fig2}H) that precisely conserves the desired correlation structure between neurons:
    \begin{equation}
        w_{ij}' = g^{-1}[\CC(S_i, S_j)] \approx \frac{\CC(S_i, S_j) - g_0}{g_1} \,,
        \label{eqn:cc-w}
    \end{equation}
    with constants $g_0$ and $g_1$ (\cref{fig:fig2}I).
    Therefore, when an SSN learns a target distribution from data, background cross-correlations are equivalent to an offset in the initial network parameters and are automatically compensated during training.
\input{tex_files/fig3.tex}

    For now, we can conclude that the activity of SSNs constitutes a sufficient source of stochasticity for other SSNs, since all effects that follow from replacing Poisson noise in an SSN with functional output from other SSNs (which at this point still receive explicit noise) can be compensated by appropriate parameter adjustments.
    These are important preliminary conclusions for the next sections, where we show how all noise can be eliminated in an ensemble of interconnected SSNs endowed with synaptic plasticity without significant penalty to their respective functional performance.

%% file: tex_files/fig2.tex
\begin{figure*}[!ht]
    \centering
    \includegraphics[width=1.\linewidth]{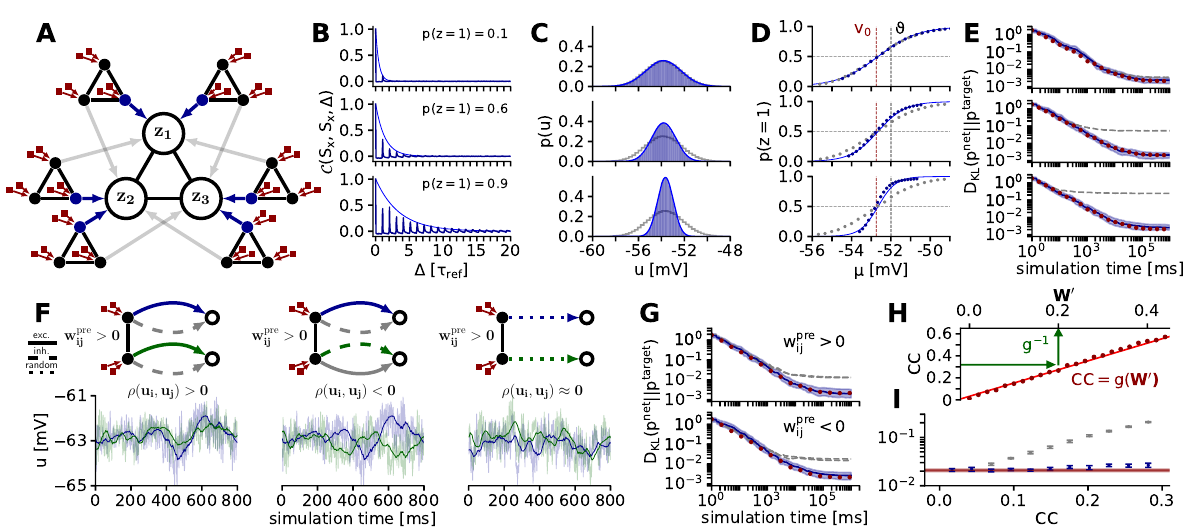}
    \caption{
        Effect of correlated background on SSN dynamics and compensation through reparametrization.
        \textbf{(A)} Feedforward replacement of Poisson noise by spiking activity from other SSNs.
                     In this illustration, the principal SSN consists of three neurons receiving background input only from other functional SSNs that sample from their own predetermined target distribution.
                     For clarity, only two out of a total of [260, 50, 34] (top to bottom in (B)) background SSNs per neuron are shown here.
                     By modifying the background connectivity (gray and blue arrows) the amount of cross-correlation in the background input can be controlled.
                     At this stage, the background SSNs still receive Poisson input (red boxes).
        \textbf{(B)} By appropriate parametrization of the background SSNs, we adjust the mean spike frequency of the background neurons (blue) to study the effect of background autocorrelations $\corr (S_x, S_x, \Delta)$.
                     Higher firing probabilities increase the chance of evoking bursts, which induce background autocorrelations for the neurons in the principal SSN at multiples of $\tauref$ (dark blue: simulation results; light blue: $e^{k \ln{\bar{p}}}$ with $k = \frac{\Delta}{\tauref}$, see \cref{eqn:acf}).
        \textbf{(C)} Background autocorrelation narrows the FMP distribution of neurons in the principal SSN: simulation (blue bars) and the theoretical prediction (\cref{eqn:fmp}, blue line) vs. background Poisson noise of the same rate (gray).
                     Background intensities correspond to (B).
        \textbf{(D)} Single-neuron activation functions corresponding to (B,C) and the theoretical prediction (\cref{eqn:actfct-approx}, blue line).
                     For autocorrelated noise, the slope of the response curve changes, but the inflection point (with $p(z=1) = 0.5$) is conserved.
        \textbf{(E)} Kullback-Leibler divergence $\DKL{\pnet}{\ptarget}$ (median and range between the first and third quartile) for the three cases shown in (B,C,D) after sampling from 50 different target distributions with 10 different random seeds for the 3-neuron network depicted in (A).
                     Appropriate reparametrization can fully cancel out the effect of background autocorrelations (blue).
                     The according results without reparametrization (gray) and with Poisson input (red) are also shown.
        \textbf{(F)} A pair of interconnected neurons in a background SSN generates correlated noise, as given by \cref{eqn:cc}.
                     The effect of cross-correlated background on a pair of target neurons depends on the nature of synaptic projections from the background to the principal SSN.
                     Here, we depict the case where their interaction $\mathrm{w}^\mathrm{pre}_\mathrm{ij}$ is excitatory; the inhibitory case is a mirror image thereof.
                     Left: If forward projections are of the same type, postsynaptic potentials will be positively correlated.
                     Middle: Different synapse types in the forward projection only change the sign of the postsynaptic potential correlations.
                     Right: For many background inputs with mixed connectivity patterns, correlations can average out to zero even when all input correlations have the same sign.
        \textbf{(G)} Same experiment as in (E), with background connection statistics adjusted to compensate for input cross-correlations.
                     The uncompensated cases from (F, left) and (F, middle) are shown in gray.
        \textbf{(H)} Correlation-cancelling reparametrization in the principal SSN.
                     By transforming the state space from $\bs z \in \{0, 1\}^n$ to $\bs z' \in \{-1, 1\}$, input correlations attain the same functional effect as synaptic weights (\cref{eqn:cc-w}); simulation results given as red dots, linear fit as red line.
                     Weight rescaling followed by a transformation back into the $\bs z \in \{0, 1\}^n$ state space, shown in green (which affects both weights and biases) can therefore alleviate the effects of correlated background.
        \textbf{(I)} Similar experiment as in (E) for a network with ten neurons, with parameters adjusted to compensate for input cross-correlations.
                     As in the case of autocorrelated background, cross-correlations can be cancelled out by appropriate reparametrization.
                     }
    \label{fig:fig2}
    \end{figure*}

%% file: tex_files/fig3.tex
\begin{figure*}[ht]
\centering
\includegraphics[width=1.\linewidth]{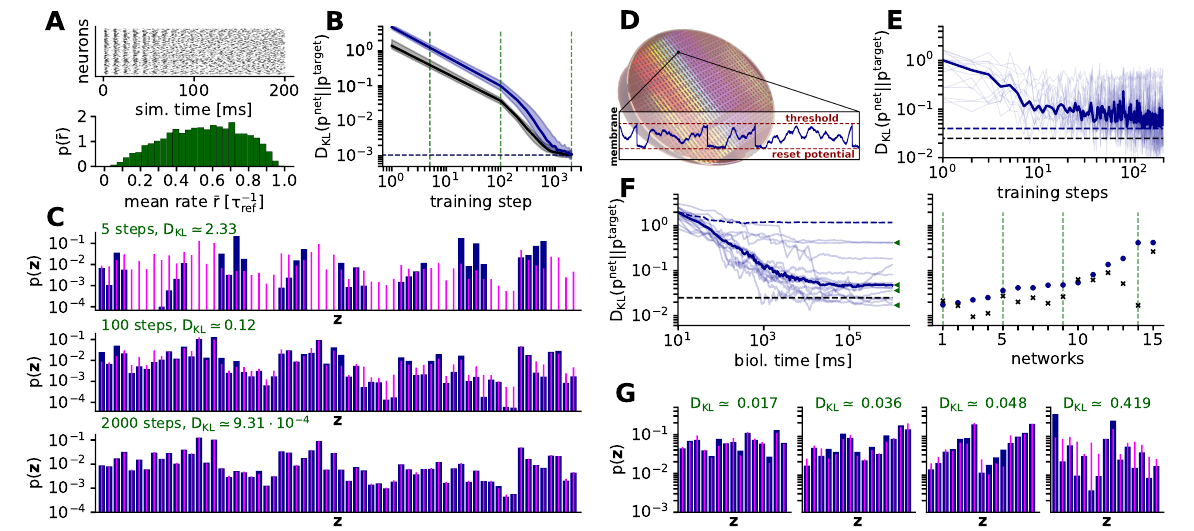}
\caption{
    Sampling without explicit noise from a set of predefined target distributions in software (A-C) and on a neuromorphic substrate (D-G).
    \textbf{(A)} (top) Temporal evolution of spiking activity in an ensemble of 100 interconnected 6-neuron SSNs with no source of explicit noise.
                 An initial burst of regular activity caused by neurons with a strong enough positive bias quickly transitions to asynchronous irregular activity due to inhibitory synapses.
                 (bottom) Distribution of mean neuronal firing rates of the ensembles shown in (B,C) after training.
    \textbf{(B)} Median sampling quality of the above ensemble during learning, for a test sampling period of $10^6$ms.
                 At the end of the learning phase, the sampling quality of individual networks in the ensemble (blue) is on par with the one obtained in the theoretically ideal case of independent networks with Poisson background (black).
                 Error bars given over 5 simulation runs with different random seeds.
    \textbf{(C)} Illustration of a single target distribution (magenta) and corresponding sampled distribution (blue) of a network in the ensemble at several stages of the learning process (dashed green lines in (B)).
    \textbf{(D)} Photograph of a wafer from the BrainScaleS neuromorphic system used in (E), (F) and (G) before post-processing (i.e., adding additional structures like buses on top), which would mask the underlying modular structure.
                 Blue: exemplary membrane trace of an analog neuron receiving Poisson noise.
    \textbf{(E)} Performance of an ensemble consisting of 15 4-neuron SSNs with no external noise during learning on the neuromorphic substrate, shown in light blue for each SSN and with the median shown in dark blue.
                 The large fluctuations compared to (B) are a signature of the natural variability of the substrate's analog components.
                 The dashed blue line represents the best achieved median performance at $\DKL{\pnet}{\ptarget} = 3.99 \times 10^{-2}$.
                 For comparison, we also plot the optimal median performance for the theoretically ideal case of independent, Poisson-driven SSNs emulated on the same substrate, which lies at $\DKL{\pnet}{\ptarget} = 2.49 \times 10^{-2}$ (dashed black line).
    \textbf{(F)} Left: Demonstration of sampling in the neuromorphic ensemble of SSNs after 200 training steps.
                 Individual networks in light blue, median performance in dark blue.
                 Dashed blue line: median performance before training.
                 Dashed black line: median performance of ideal networks, as in (E).
                 Right: Best achieved performance, after \SI{100}{s} of bio time (\SI{10}{ms} of hardware time) for all SSNs in the ensemble depicted as blue dots (sorted from lowest to highest $\DKLsolo$).
                 For comparison, the same is plotted as black crosses for their ideal counterparts.
    \textbf{(G)} Sampled (blue) and target (magenta) distributions of four of the 15 SSNs.
                 The selection is marked in (F) with green triangles (left) and vertical green dashed lines (right).
                 Since we made no particular selection of hardware neurons according to their behavior, hardware defects have a significant impact on a small subset of the SSNs. 
                 Despite these imperfections, a majority of SSNs perform close to the best value permitted by the limited weight resolution (4 bits) of the substrate.
    }
\label{fig:fig3}
\end{figure*}

%% file: tex_files/sampling_without_noise.tex
\subsection{Sampling without explicit noise in large ensembles}

We initialized an ensemble of 100 6-neuron SSNs with an inter-network connectivity of $\epsilon=0.1$ and random synaptic weights.
As opposed to the previous experiments, none of the neurons in the ensemble receive explicit Poisson input and the activity of the ensemble itself acts as a source of stochasticity instead, as depicted in \cref{fig:fig1}D.
No external input is needed to kick-start network activity, as some neurons spike spontaneously, due to the random initialization of parameters (see \cref{fig:fig3}A).
The existence of inhibitory weights disrupts the initial regularity, initiating the sampling process.
Ongoing learning (\Cref{eq:CDw,eq:CDb}) shapes the sampled distributions towards their respective targets (\cref{fig:fig3}B), the parameters of which were drawn randomly (see Supporting information).
Our ensemble achieved a sampling performance (median $\DKLsolo$) of $1.06^{+0.27}_{-0.40}\times 10^{-3}$, which is similar to the median performance of an idealized setup (independent, Poisson-driven SSNs as in \cite{petrovici2016stochastic}) of $1.05^{+0.15}_{-0.35}\times 10^{-3}$ (errors are given by the first and third quartile).
To put the above $\DKLsolo$ values in perspective, we compare the sampled and target distributions of one of the SSNs in the ensemble at various stages of learning (\cref{fig:fig3}C).
Thus, despite the fully deterministic nature of the system, the network dynamics and achieved performance after training is essentially indistinguishable from that of networks harnessing explicit noise for the representation of probability.
Instead of training ensembles, they can also be set up by translating the parameters of the target distributions to neurosynaptic parameters directly, as discussed in the previous section (see Supporting information, \cref{fig:figS2}).

%% file: tex_files/neuromorphic.tex
\subsection{Implementation on a neuromorphic substrate}

To test the robustness of our results, we studied an implementation of noise-free sampling on an artificial neural substrate, which incorporates unreliable components and is therefore significantly more difficult to control.
For this, we used the BrainScaleS system \cite{schemmel2008wafer}, a mixed-signal neuromorphic platform with analog neurosynaptic dynamics and digital inter-neuron communication (\cref{fig:fig3}D, see also Supporting information, \cref{fig:figS3}).
A major advantage of this implementation is the emulation speedup of $10^4$ with respect to biological real-time; however, for clarity, we shall continue using biological time units instead of actual emulation time.

The additional challenge for our neuronal ensemble is to cope with the natural variability of the substrate, caused mainly by fixed-pattern noise, or with other limitations such as a finite weight resolution (4 bits) or spike loss, which can all be substantial \cite{petrovici2014characterization, schmitt2017neuromorphic}.
It is important to note that the ability to function when embedded in an imperfect substrate with significant deviations from an idealized model represents a necessary prerequisite for viable theories of biological neural function.

We emulated an ensemble of 15 4-neuron SSNs, with an inter-SSN connectivity of $\epsilon = 0.2$ and with randomly drawn target distributions (see Supporting information).
The biases were provided by additional bias neurons and adjusted during learning via the synaptic weights between bias and sampling neurons, along with the synapses within the SSNs, using the same learning rule as before (\Cref{eq:CDw,eq:CDb}).
After 200 training steps, the ensemble reached a median $\DKLsolo$ of $3.99^{+1.27}_{-1.15} \cdot 10^{-2}$ (errors given by the distance to the first and third quartile) compared to $1.18^{+0.47}_{-0.55}$ before training (\cref{fig:fig3}E).
As a point of reference, we also considered the idealized case by training the same set of SSNs without interconnections and with every neuron receiving external Poisson noise generated from the host computer, reaching a $D_\mathrm{KL}$ of $2.49^{+3.18}_{-0.71}\cdot 10^{-2}$.

This relatively small performance loss of the noise-free ensemble compared to the ideal case confirms the theoretical predictions and simulation results.
Importantly, this was achieved with only a rather small ensemble, demonstrating that large numbers of neurons are not needed for realizing this computational paradigm.

In \cref{fig:fig3}F\movieref{1}, we show the sampling dynamics of all emulated SSNs after learning.
While most SSNs are able to approximate their target distributions well, some sampled distributions are significantly skewed (\cref{fig:fig3}G).
This is caused by a small subset of dysfunctional neurons, which we have not discarded beforehand, in order to avoid an implausibly fine-tuned use-case of the neuromorphic substrate.
These effects become less significant in larger networks trained on data instead of predefined distributions, where learning can naturally cope with such outliers by assigning them smaller output weights.
Nevertheless, these results demonstrate the feasibility of self-sustained Bayesian computation through sampling in physical neural substrates, without the need for any source of explicit noise.
Importantly, and in contrast to other approaches \cite{jordan2017stochastic}, every neuron in the ensemble plays a functional role, with no neuronal real-estate being dedicated to the production of (pseudo-)randomness.

%% file: tex_files/applications.tex
\subsection{Ensembles of hierarchical SSNs}
When endowed with appropriate learning rules, hierarchical spiking networks can be efficiently trained on high-dimensional visual data \cite{petrovici2017pattern, schmitt2017neuromorphic, leng2018spiking, lee2016training, zenke2018superspike, kheradpisheh2018stdp}.
Such hierarchical networks are characterized by the presence of several layers, with connections between consecutive layers, but no lateral connections within the layers themselves.
When both feedforward and feedback connections are present, such networks are able to both classify and generate images that are similar to those used during training.

\input{tex_files/fig4.tex}
In these networks, information processing in both directions is Bayesian in nature.
Bottom-up propagation of information enables an estimation of the conditional probability of a particular label to fit the input data.
Additionally, top-down propagation of neural activity allows generating a subset of patterns in the visible layer conditioned on incomplete or partially occluded visual stimulus.
When no input is presented, such networks will produce patterns similar to those enforced during training ("dreaming").
In general, the exploration of a multimodal solution space in generative models is facilitated by some noise-generating mechanism.
We demonstrate how even a small interconnected set of hierarchical SSNs can perform these computations self-sufficiently, without any source of explicit noise.

We used an ensemble of four 3-layer hierarchical SSNs trained on a subset of the EMNIST dataset \cite{cohen2017emnist}, an extended version of the widely used MNIST dataset \cite{lecun1998gradient} that includes digits as well as capital and lower-case letters.
All SSNs had the same structure, with 784 visible units, 200 hidden units and 5 label units (\cref{fig:fig4}A).
To emulate the presence of networks with different functionality, we trained each of them on a separate subset of the data.
(To combine sampling in space with sampling in time, multiple networks can also be trained on the same data, see Supporting information \cref{fig:figS5} \movieref{2}.)
Since training the spiking ensemble directly was computationally prohibitive, we trained four Boltzmann machines on the respective datasets and then translated the resulting parameters to neurosynaptic parameters of the ensemble using the methods described earlier in the manuscript (see also Supporting information, \cref{fig:figS2}).

To test the discriminative properties of the SSNs in the ensemble, one was stimulated with visual input, while the remaining three were left to freely sample from their underlying distribution.
We measured a median classification rate of $91.5^{+3.6}_{-3.0}\%$ with errors given by the distance to the first and third quartile, which is close to the $94.0^{+2.1}_{-1.5}\%$ achieved by the idealized reference setup provided by the abstract Boltzmann machines (\cref{fig:fig4}B).
At the same time, all other SSNs remained capable of generating recognizable images (\cref{fig:fig4}C).
It is expected that direct training and a larger number of SSNs in the ensemble would further improve the results, but a functioning translation from the abstract to the biological domain already underpins the soundness of the underlying theory.

Without visual stimulus, all SSNs sampled freely, generating images similar to those on which they were trained (\cref{fig:fig4}D).
Without any source of explicit noise, the SSNs were capable to mix between the relevant modes (images belonging to all classes) of their respective underlying distributions, which is a hallmark of a good generative model.
We further extended these results to an ensemble trained on the full MNIST dataset, reaching a similar generative performance for all networks (see Supporting information \cref{fig:figS5} \movieref{2}).

To test the pattern completion capabilities of the SSNs in the ensemble, we stimulated them with incomplete and ambiguous visual data (\cref{fig:fig4}E).
Under these conditions, SSNs only produced images compatible with the stimulus, alternating between different image classes, in a display of pattern rivalry.
As in the case of free dreaming, the key mechanism facilitating this form of exploration was provided by the functional activity of other neurons in the ensemble.

%% file: tex_files/fig4.tex
\begin{figure*}[!th]
\centering
\includegraphics[width=1.\linewidth]{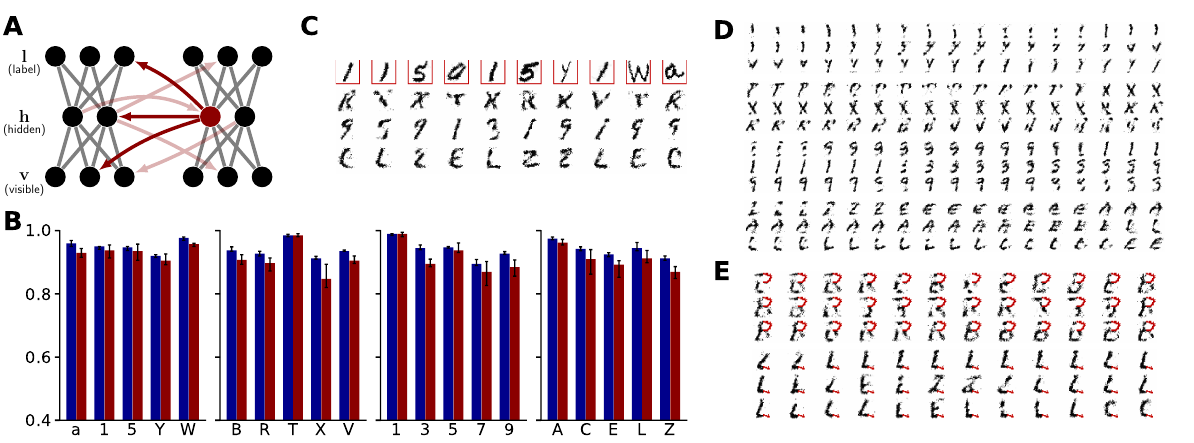}
\caption{
    Bayesian inference on visual input.
    \textbf{(A)} Illustration of the connectivity between two hierarchical SSNs in the simulated ensemble.
                 Each SSN had a visible layer \textbf{v}, a hidden \textbf{h} and a label layer \textbf{l}.
                 Neurons in the same layer of an SSN were not interconnected.
                 Each neuron in an SSN received only activity from the hidden layers of other SSNs as background (no sources of explicit noise).
    \textbf{(B)} An ensemble of four such SSNs (red) was trained to perform generative and discriminative tasks on visual data from the EMNIST dataset.
                 We used the classification rate of restricted Boltzmann machines trained with the same hyperparameters as a benchmark (blue).
                 Error bars are given (on blue) over 10 test runs and (on red) over 10 ensemble realizations with different random seeds.
    \textbf{(C)} Illustration of a scenario where one of the four SSNs (red boxes) received visual input for classification (B).
                 At the same time, the other SSNs continuously generated images from their respective learned distributions.
    \textbf{(D)} Pattern generation and mixing during unconstrained dreaming.
                 Here, we show the activity of the visible layer of all four networks from (B), each spanning three rows.
                 Time evolves from left to right.
                 For further illustrations of the sampling process in the ensemble of hierarchical SSNs, see Supporting information, \cref{fig:figS4,fig:figS5} \movieref{2}.
    \textbf{(E)} Pattern completion and rivalry for two instances of incomplete visual stimulus.
                 The stimulus consisted of the top right and bottom right quadrant of the visible layer, respectively.
                 In the first run, we clamped the top arc of a ``B'' compatible with either a ``B'' or an ``R'' (top three rows, red), in the second run we chose the bottom line of an ``L'' compatible with an ``L'', an ``E'', a ``Z'' or a ``C'' (bottom three rows, red).
                 An ensemble of SSNs performs Bayesian inference by implicitly evaluating the conditional distribution of the unstimulated visible neurons, which manifests itself here as sampling from all image classes compatible with the ambiguous stimulus (see also Supporting information, \cref{fig:figS6}).
    }
\label{fig:fig4}
\end{figure*}

%% file: tex_files/discussion.tex
\section{Discussion}

Based on our findings, we argue that sampling-based Bayesian computation can be implemented in fully deterministic ensembles of spiking networks without requiring any explicit noise-generating mechanism.
Our approach has a firm theoretical foundation in the theory of sampling spiking neural networks, upon which we formulate a rigorous analysis of network dynamics and learning in the presence or absence of noise.

While in biology various explicit sources of noise exist \cite{faisal2008noise, branco2009probability, white2000channel}, these forms of stochasticity are either too weak (in case of ion channels) or too high-dimensional for efficient exploration (in the case of stochastic synaptic transmission, as used for, e.g., reinforcement learning \cite{seung2003learning}).
Furthermore, a rigorous mathematical framework for neural sampling with stochastic synapses is still lacking.
On the other hand, in the case of population codes, neuronal population noise can be highly correlated, affecting information processing by, e.g., inducing systematic sampling biases \cite{averbeck2006neural}.

In our proposed framework, each network in an ensemble plays a dual role: while fulfilling its assigned function within its home subnetwork, it also provides its peers with the spiking background necessary for stochastic search within their respective solution spaces.
This enables a self-consistent and parsimonious implementation of neural sampling, by allowing all neurons to take on a functional role and not dedicating any resources purely to the production of background stochasticity.
The underlying idea lies in adapting neuro-synaptic parameters by (contrastive) Hebbian learning to compensate for auto- and cross-correlations induced by interactions between the functional networks in the ensemble.
Importantly, we show that this does not rely on the presence of a large number of independent presynaptic partners for each neuron, as often assumed by models of cortical computation that use Poisson noise (see, e.g., \cite{xie2004learning}).
Instead, only a small number of ensembles is necessary to implement noise-free Bayesian sampling.
This becomes particularly relevant for the development of neuromorphic platforms by eliminating the computational footprint imposed by the generation and distribution of explicit noise, thereby reducing power consumption and bandwidth constraints.

For simplicity, we chose networks of similar size in our simulations.
However, the presented results are not contingent on network sizes in the ensemble and largely independent of the particular functionality (underlying distribution) of each SSN.
Their applicability to scenarios where different SSNs learn to represent different data is particularly relevant for cortical computation, where weakly interconnected areas or modules are responsible for distinct functions \cite{chen2008revealing,bullmore2009complex,meunier2010modular,bertolero2015modular,song2005highly}.
Importantly, these ideas scale naturally to larger ensembles and larger SSNs.
Since each neuron only needs a small number of presynaptic partners from the ensemble, larger networks lead to a sparser interconnectivity between SSNs in the ensemble and hence soften structural constraints.
Preliminary simulations show that the principle of using functional output as noise can even be applied to connections \textit{within} a single SSN, eliminating the artificial separation between network and ensemble connections (see \cref{fig:figS7} and Video S3 in the Supporting information).

Even though we have used a simplified neuron model in our simulations to reduce computation time and facilitate the mathematical analysis, we expect the core underlying principles to generalize. 
This is evidenced by our results on neuromorphic hardware, where the dynamics of individual neurons and synapses differ significantly from the mathematical model.
Such an ability to compute with unreliable components represents a particularly appealing feature in the context of both biology and emerging nanoscale technologies.

Finally, the suggested noise-free Bayesian brain reconciles the debate on spatial versus temporal sampling \cite{ma2006bayesian, orban2016neural}.
In fact, the networks of spiking neurons that provide each other with virtual noise may be arranged in parallel sensory streams.
An ambiguous stimulus will trigger different representations on each level of these streams, forming a hierarchy of probabilistic population codes.
While these population codes learn to cover the full sensory distribution in space, they will also generate samples of the sensory distribution in time (see \cref{fig:figS5} in the Supporting information).
Attention may select the most likely representation, while suppressing the representations in the other streams.
Analogously, possible actions may be represented in parallel motor streams during planning and a motor decision may select the one to be performed.
When recording in premotor cortex, such a selection causes a noise reduction \cite{churchland2006neural}, that we suggest is effectively the signature of choosing the most probable action in a Bayesian sense.

%% file: tex_files/conclusion.tex
\section{Conclusion}
From a generic Bayesian perspective, cortical networks can be viewed as generators of target distributions. 
To enable such computation, models assume neurons to possess sources of perfect, well-behaved noise -- an assumption that is both
impractical and at odds with biology. 
We showed how local plasticity in an ensemble of spiking networks allows them to co-shape their activity towards a set of well-defined targets, while reciprocally using the very same activity as a source of (pseudo-)stochasticity. 
This enables purely deterministic networks to simultaneously learn a variety of tasks, completely removing the need for true randomness. 
While reconciling the sampling hypothesis with the deterministic nature of single neurons, this also offers an efficient blueprint for in-silico implementations of sampling-based inference.

%% file: supplement/calc_details.tex
\section{Calculations}

\subsection{Free membrane potential distribution with colored noise}

In the high-conductance state (HCS), it can be shown that the temporal evolution of the free membrane potential (FMP) of an LIF neuron stimulated by balanced Poisson inputs is equivalent to an Ornstein-Uhlenbeck (OU) process with the following Green's function \cite{petrovici2016form}:
\begin{align}
f(u,t & |u_0) = \sqrt{\frac{1}{2 \pi \sigma^2 (1-e^{-2\theta t})}} \nonumber \\ 
\cdot &\exp{\bigg(- \frac{1}{2 \sigma^2}\frac{(u-\mu+(\mu - u_0)e^{-\theta t})^2}{1-e^{-2\theta t}}\bigg)}\,. \label{eq:SI1}
\end{align}
with
\begin{subequations}
\begin{align}
\theta &= \frac{1}{\tau^\mathrm{syn}}\,, \\
\mu &= \frac{\gl \El + \sum_{k\in\{\mathrm{e},\mathrm{i}\}} \nu_k w_k \Erev_k \tau^\mathrm{syn}}{\braket{\gtot}}\,, \label{eq:SI_mu}\\
\sigma^2 &= \frac{\sum_{k\in\{\mathrm{e},\mathrm{i}\}} \nu_k w_k^2 (\Erev_k - \mu)^2 \tau^\mathrm{syn}}{\braket{\gtot}^2}\,,\\
\braket{\gtot} &=\ \gl + \sum_{k\in\{\mathrm{e},\mathrm{i}\}} w_k \nu_k \tausyn_k\,,
\end{align}
\end{subequations}
where $\nu_k$ are the noise frequencies, $w_k$ the noise weights and we dropped the index notation $u_k^\mathrm{free}$ used in previous sections for convenience.
The stationary FMP distribution is then given by a Gaussian \cite{gerstner2002spiking, petrovici2016form}:
\begin{equation}
f(u) = \sqrt{\frac{1}{2 \pi \sigma^2}} \exp{\bigg(- \frac{(u-\mu)^2}{2 \sigma^2}\bigg)}\,. \label{eq:SI1_2}
\end{equation}
Replacing the white noise $\eta(t)$ in the OU process, defined by $\braket{\eta} = \mathrm{const.}$ and $\braket{\eta(t)\eta(t')} = \nu \delta(t-t') + \nu^2$ \cite{gerstner2002spiking}, with (Gaussian) colored noise $\eta_c$, defined by $\braket{\eta_c} = \mathrm{const.}$ and $\braket{\eta_c(t)\eta_c(t')} = \gamma(t-t')$ \cite{haunggi1994colored} where $\gamma(t-t')$ is a function that does not vanish for $t-t' \neq 0$, the stationary solution of the FMP distribution is still given by a Gaussian with mean $\mu'$ and width $\sigma'$ \cite{haunggi1994colored,caceres1999harmonic}.
Since the noise correlations only appear when calculating higher-order moments of the FMP, the mean value of the FMP distribution remains unchanged $\mu' = \mu$.
However, the variance $\sigma'^2 = \braket{\big(u(t)-\braket{u(t)}\big)^2}$ of the stationary FMP distribution changes due to the correlations, as discussed in the next section.

\subsection{Width of free membrane potential distribution}

In the HCS, the FMP can be approximated analytically as \cite{petrovici2016form}
\begin{align}
   u(t) &= u_0 + \sum_{k\in\{\mathrm{e},\mathrm{i}\}} \sum_{\mathrm{spikes\hspace{.75mm} s}} \Lambda_k \Theta(t-t_s) \nonumber\\ 
   \cdot &\bigg[\exp\bigg(-\frac{t-t_s}{\tausyn_k}\bigg)-\exp\bigg(-\frac{t-t_s}{\braket{\taueff}}\bigg) \bigg]\,,  \label{eq:SI_PSP}
\end{align}
with
\begin{subequations}
\begin{align}
   u_0 =\ &\frac{\gl\El + (\braket{\gtot}-\gl)\mu}{\braket{\gtot}}\,, \\
   \Lambda_k =\ &\frac{\tausyn_k w_k \big(\Erev_k- \mu\big)}{\braket{\gtot}\big(\tausyn_k-\braket{\taueff}\big)}\,, \label{eq:SI_lambda}\\
   \braket{\taueff} =\ &\frac{\Cm}{\braket{\gtot}} \label{eq:SI_taueff}\,.
\end{align}
\end{subequations}
By explicitly writing the excitatory and inhibitory noise spike trains as $S_\mathrm{e/i}(t') = \sum_\mathrm{spikes\ s} \delta(t' - t_s)$, this can be rewritten to
\begin{subequations}
\begin{align}
u(t) &= u_0 + \sum_{k\in\{\mathrm{e},\mathrm{i}\}} \Lambda_k \int \mathrm{d}t' S_k(t') \Theta(t-t') \nonumber \\
\cdot &\bigg[\exp\bigg(-\frac{t-t'}{\tausyn_k}\bigg)-\exp\bigg(-\frac{t-t'}{\braket{\taueff}}\bigg) \bigg] \label{eq:SI_FMP} \\
&= u_0 + \Lambda_\mathrm{e} \big(S_\mathrm{e} * \kappa_e\big)(t) + \Lambda_\mathrm{i} \big(S_\mathrm{i} * \kappa_i\big)(t) \\
&= u_0 + \big[ (\Lambda_\mathrm{e} S_\mathrm{e} + \Lambda_\mathrm{i} S_\mathrm{i}) * \kappa\big](t)\,, \label{eq:eq1}
\end{align}
\end{subequations}
where $*$ denotes the convolution operator and with
\begin{align}
\kappa_\mathrm{e/i}(t) &= \Theta(t)\bigg[\exp\bigg(-\frac{t}{\tau_\mathrm{e/i}^{\mathrm{syn}}}\bigg)-\exp\bigg(-\frac{t}{\braket{\taueff}}\bigg) \bigg] \,.
\end{align}
For simplicity, we assume $\tau_\mathrm{e}^\mathrm{syn} = \tau_\mathrm{i}^\mathrm{syn}$.
The width of the FMP distribution can now be calculated as
\begin{subequations}
\begin{align}
&\braket{\big(u(t)-\braket{u(t)}\big)^2} = \braket{u(t)^2} - \braket{u(t)}^2 \label{eq:SI_FMP2}\\
&= \braket{\big[ u_0 + (S_\mathrm{tot} * \kappa)(t)\big]^2} - \braket{u_0 + (S_\mathrm{tot} * \kappa)(t)}^2 \\
&= \braket{\big[ (S_\mathrm{tot} * \kappa)(t)\big]^2} - \braket{(S_\mathrm{tot} * \kappa)(t)}^2 \\
&= \braket{\big[ (S_\mathrm{tot} * \kappa)(t) - \braket{(S_\mathrm{tot} * \kappa)(t)}\big]^2} \, ,
\end{align}
\end{subequations}
where the average is calculated over $t$ and $S_\mathrm{tot}(t) = \Lambda_\mathrm{e} S_\mathrm{e}(t) + \Lambda_\mathrm{i} S_\mathrm{i}(t)$. 
Since the average is an integral over $t$, i.e. $\braket{(\cdot)} \rightarrow \lim\limits_{T \rightarrow \infty} \frac{1}{T} \int_{-T/2}^{T/2} (\cdot)\ \mathrm{d}t$, we can use the identity $\int (f*g)(t)\ \mathrm{d}t = (\int f(t)\ \mathrm{d}t)(\int g(t)\ \mathrm{d}t)$, that is $\braket{f*g} = \braket{f}\int g(t)\ \mathrm{d}t = \braket{f}*g$ in the limit of $T \rightarrow \infty$, to arrive at the following solution:
\begin{align}
&\braket{\big(u(t)-\braket{u(t)}\big)^2} \nonumber \\ &=\braket{\big[(S_\mathrm{tot}(t) - \braket{S_\mathrm{tot}(t)})*\kappa(t)\big]^2}\,.
\end{align}
More generally, we obtain with a similar calculation the autocorrelation function (ACF) of the FMP:
\begin{align}
&\braket{\bar{u}(t)\bar{u}(t+\Delta)} \nonumber \\ 
&= \braket{\big((\bar{S}_\mathrm{tot}*\kappa)(t)\big)\big((\bar{S}_\mathrm{tot}*\kappa)(t+\Delta)\big)}\,,
\end{align}
with $\bar{x}(t) = x(t) - \braket{x(t)}$ and by using $\braket{\bar{u}(t)\bar{u}(t+\Delta)} = \braket{u(t)u(t+\Delta)} - \braket{u(t)}^2$.
This can be further simplified by applying the Wiener--Khintchine theorem \cite{wiener1930generalized, khintchine1934korrelationstheorie}, which states that $\lim\limits_{T \rightarrow \infty}\braket{x(t)x(t+\Delta)}_T = \mathcal{F}^{-1}\big(|\mathcal{F}(x)|^2)\big)(\Delta)$ with $\braket{(\cdot)}_T \rightarrow \int_{-T/2}^{T/2} (\cdot)\ \mathrm{d}t$ (due to $\int x(t)x(t+\Delta)\ \mathrm{d}t = \big(x(t)*x(-t)\big)(\Delta)$). Thus, for the limit $T \rightarrow \infty$, we can rewrite this as
\begin{subequations}
\begin{align}
&\braket{\big((\bar{S}_\mathrm{tot}*\kappa)(t)\big)\big((\bar{S}_\mathrm{tot}*\kappa)(t+\Delta)\big)} \nonumber \\
&= \lim\limits_{T \rightarrow \infty} \frac{1}{T} \mathcal{F}^{-1}\big(|\mathcal{F}(\bar{S}_\mathrm{tot}*\kappa)|^2\big)(\Delta)\\
&= \lim\limits_{T \rightarrow \infty} \frac{1}{T} \mathcal{F}^{-1}\big(|\mathcal{F}(\bar{S}_\mathrm{tot})\mathcal{F}(\kappa)|^2\big)(\Delta) \\
&= \lim\limits_{T \rightarrow \infty} \frac{1}{T} \bigg(\mathcal{F}^{-1}\big(|\mathcal{F}(\bar{S}_\mathrm{tot})|^2\big) \nonumber \\
&\ \ *\mathcal{F}^{-1}\big(|\mathcal{F}(\kappa)|^2\big)\bigg)(\Delta)\,,
\end{align}
\end{subequations}
and by applying the Wiener--Khintchine theorem again in reverse
\begin{align}
\braket{\bar{u}(t)\bar{u}(t+\Delta)} &= \big( \lim\limits_{T \rightarrow \infty} \frac{1}{T} \braket{\big[\bar{S}_\mathrm{tot}(t)\big]\big[\bar{S}_\mathrm{tot}(t+\Delta')\big]}_T \nonumber \\
&*\braket{\big[\kappa(t)\big]\big[\kappa(t+\Delta')\big]}_\infty\big)(\Delta)\,, \label{eq:SI_lim}
\end{align}
where the variance of the FMP distribution is given for $\Delta = 0$.
Thus, the unnormalized ACF of the FMP can be calculated by convolving the unnormalized ACF of the background spike trains ($S_\mathrm{tot}$) and the PSP shape ($\kappa$). 
In case of independent excitatory and inhibitory Poisson noise (i.e., $\braket{\bar{S}(t)\bar{S}(t')} = \nu \delta(t-t')$), we get
\begin{subequations}
\begin{align}
\braket{\big[\bar{S}_\mathrm{tot}(t)\big]\big[\bar{S}_\mathrm{tot}(t+\Delta')\big]} &= \Lambda_\mathrm{e}^2 \braket{\bar{S}_\mathrm{e}(t)\bar{S}_\mathrm{e}(t+\Delta')} \nonumber \\ 
&\ \ + \Lambda_\mathrm{i}^2 \braket{\bar{S}_\mathrm{i}(t)\bar{S}_\mathrm{i}(t+\Delta')}\\
&= \sum_{k\in\{\mathrm{e},\mathrm{i}\}} \Lambda_k^2 \nu_k \delta(\Delta')
\end{align}
\end{subequations}
and therefore
\begin{subequations}
\begin{align}
\mathrm{Var}(u) &= \big(\sum_{k\in\{\mathrm{e},\mathrm{i}\}} \Lambda_k^2 \nu_k \delta(\Delta')\nonumber \\
&\ \ *\braket{\big[\kappa(t)\big]\big[\kappa(t+\Delta')\big]}_\infty\big)(\Delta=0)\\
&= \sum_{k\in\{\mathrm{e},\mathrm{i}\}} \Lambda_k^2 \nu_k \braket{\kappa^2(t)} \\
&= \sum_{k\in\{\mathrm{e},\mathrm{i}\}} \Lambda_k^2 \nu_k \int_0^\infty \kappa^2(t)\ \mathrm{d}t\,,
\end{align}
\end{subequations}
which agrees with the result given in \cite{petrovici2016form}.
If the noise spike trains are generated by processes with refractory periods, the absence of spikes between refractory periods leads to negative contributions in the ACF of the noise spike trains.
This leads to a reduced value of the variance of the FMP and hence, also to a reduced width of the FMP distribution.
The factor $\sqrt{\beta}$ by which the width of the FMP distribution (Eqn.~(11)) changes due to the introduction of colored background noise is given by
\begin{align}
\beta &= \frac{\sigma^2_\mathrm{colored}}{\sigma^2_\mathrm{Poisson}} \label{eq:SI_width_rescale} \\
&= \frac{\int \mathrm{d}\Delta\ \braket{\bar{S}_\mathrm{tot}(t)\bar{S}_\mathrm{tot}(t+\Delta)}\cdot \int \mathrm{d}t\ \kappa(t)\kappa(t+\Delta)}{\sum_{k\in\{\mathrm{e},\mathrm{i}\}} \Lambda_k^2 \nu_k \int_0^\infty \kappa^2(t)\ \mathrm{d}t} \,.
\end{align}
For the simplified case of a Poisson process with refractory period, one can show that $\int \mathrm{d}\Delta\ \braket{\bar{S}_\mathrm{tot}(t)\bar{S}_\mathrm{tot}(t+\Delta)}$ has a reduced value compared to a Poisson process without refractory period \cite{gerstner2002spiking}, leading to $\beta \leq 1$. Even though we do not show this here for neuron-generated spike trains, the two cases are similar enough that $\beta \leq 1$ can be assumed to apply in this case as well. 

In the next section, we will show that the factor $\beta$ can be used to rescale the inverse slope of the activation function to transform the activation function of a neuron receiving white noise to the activation function of a neuron receiving equivalent (in frequency and weights), but colored noise. That is, the rescaling of the FMP distribution width due to the autocorrelated background noise translates into a rescaled inverse slope of the activation function.

\subsection{Approximate inverse slope of LIF activation function}

As stated earlier, the FMP of an LIF neuron in the HCS is described by an OU process with a Gaussian stationary FMP distribution (both for white and colored background noise). 
As a first approximation, we can define the activation function as the probability of the neuron having a FMP above threshold (see \cref{eq:SI1_2})
\begin{align}
p(z_i = 1) &\approx \int_\vartheta^\infty f(u)\mathrm{d}u \\
&= \int_\vartheta^\infty \sqrt{\frac{1}{2 \pi \sigma^2}} \exp{\bigg(- \frac{(u-\mu)^2}{2 \sigma^2}\bigg)}\mathrm{d}u \\
&= \frac{1}{2} \left (1 - \mathrm{erf}\big(\frac{\vartheta - \mu}{\sqrt{2}\sigma} \big) \right). 
\end{align}
Even though this is only an approximation (as we are neglecting the effect of the reset), the error function is already similar to the logistic activation function observed in simulations \cite{petrovici2016stochastic}.

The inverse slope of a logistic activation function is defined at the inflection point, i.e.,
\begin{equation}
\alpha^{-1} = \frac{\mathrm{d}}{\mathrm{d}\mu}\varphi{\left(\frac{\mu - u_0}{\alpha}\right)}\bigg|_{\mu = u_0} \, .
\end{equation}
By calculating the inverse slope via the activation function derived from the FMP distribution, we get
\begin{align}
\alpha^{-1} &= \frac{\mathrm{d}}{\mathrm{d}\mu} p(z_i = 1) \bigg|_{\mu = \vartheta}\,, \label{eq:SI_inverse_slope}\\
&= \sqrt{\frac{1}{2 \pi \sigma^2}} \label{eq:SI_rescale_inv_slope} \,,
\end{align}
from which it follows that the inverse slope $\alpha$ is proportional to the width of the FMP distribution $\sigma$.
Thus, rescaling the variance of the FMP distribution by a factor $\beta$ leads, approximately, to a rescaling of the inverse slope of the activation function $\alpha' = \sqrt{\beta}\alpha$.

\subsection{Origin of side-peaks in the noise autocorrelation function}

For high rates, the spike train generated by an LIF neuron in the HCS shows regular patterns of interspike intervals which are roughly equal to the absolute refractory period.
This occurs (i) due to the refractory period introducing regularity for higher rates, since ISI's $< \tauref$ are not allowed and the maximum firing rate of the LIF neuron is bounded by $\frac{1}{\tauref}$, and (ii) due to an LIF neurons's tendency to spike consecutively when the effective membrane potential
\begin{align}
\ueff(t) &= \frac{\gl\El + \sum_{k\in\{\mathrm{e},\mathrm{i}\}} \gsyn_k(t) \Erev_k}{\gtot(t)} \,,\\
\taueff \dot{u} &= \ueff - u \,,
\end{align}
is suprathreshold after the refractory period \cite{petrovici2016form}. The probability of a consecutive spike after the neuron has spiked once at time $t$ is given by (under the assumption of the HCS)
\begin{align}
p_1 &= p(\mathrm{spike\ at\ }t+\tauref |\ \mathrm{first\ spike\ at\ }t)  \\
&= \int_\vartheta^\infty \mathrm{d}u_{t+\tauref} f(u_{t+\tauref},\ \tauref\ |\ u_t = \vartheta)\,,\nonumber
\end{align}
due to the effective membrane potential following an Ornstein-Uhlenbeck process (whereas the FMP is a low-pass filter thereof, however with a very low time constant $\taueff$), see \cref{eq:SI1}. The probability to spike again after the second refractory period is then given by
\begin{align}
p_2 &= p(\mathrm{spike\ at\ }t+2\tauref |\ \mathrm{spike\ at\ }t+\tauref,\ t) \\ 
&= \frac{\int_\vartheta^\infty \int_\vartheta^\infty \mathrm{d}u_2 \mathrm{d}u_1\ f(u_2, \tauref\ |\ u_1) f(u_1, \tauref\ |\ u_0 = \vartheta)}{\int_\vartheta^\infty \mathrm{d}u_1\ f(u_1, \tauref\ |\ u_0 = \vartheta)}\,,\nonumber
\end{align}
with $u_n = u_{t+n\tauref}$, or in general after $n-1$ spikes
\begin{align}
p_n &= p(\mathrm{spike\ at\ }t+n\tauref\ |\ \mathrm{spike\ at\ }t+(n-1)\tauref, \ldots, \ t) \nonumber \\ 
&= \int_\vartheta^\infty \mathrm{d}u_{n-1}\ f^{(n-1)}(u_{n-1}) \,, \\
f^{n}(u_n) &= \frac{\int_\vartheta^\infty \mathrm{d}u_{n-1}\ f(u_n, \tauref\ |\ u_{n-1}) f^{(n-1)}(u_{n-1})}{\int_\vartheta^\infty \mathrm{d}u_{n-1}\ f^{(n-1)}(u_{n-1})}\,, \label{eq:SIburst} 
\end{align}
for $n>1$ and $f^1(u_1)=f(u_1, \tauref\ |\ u_0 = \vartheta)$. The probability to observe $n$ spikes in such a sequence is then given by
\begin{equation}
P_n =  \prod_{i = 1}^{n-1} p_i \,,
\end{equation}
and the probability to find a burst of length $n$ (i.e., the burst ends)
\begin{equation}
p(\mathrm{burst\ of\ length\ n}) = P_n \cdot (1- p_n) \,.
\end{equation}
With this, one can calculate the average length of the occurring bursts $\sum_{i=1}^\infty i \cdot p(\mathrm{burst\ of\ length\ i})$, from which we can already see how the occurrence of bursts depends on the mean activity of the neuron.
A simple solution can be found for the special case of $\tausyn \ll \tauref$, since then the effective membrane potential distribution has already converged to the stationary distribution after every refractory period, i.e., $f(u_n, \tauref\ |\ u_{n-1}) = f(u_n)$ and hence 
\begin{align}
p_n &= p(\mathrm{spike\ at\ }t+n\tauref |\ \mathrm{spike\ at\ }t+(n-1)\tauref, \ldots, \ t) \nonumber \\ 
&= \int_\vartheta^\infty \mathrm{d}u f(u) = \bar{p}
\end{align}
for all $n$.
Thus, for this special case the average burst length can be expressed as
\begin{align}
\sum_{i=1}^\infty i \cdot p(\mathrm{burst\ of\ length\ i}) &= \sum_{i=1}^\infty i \cdot \bar{p}^{i-1}(1-\bar{p})\,, \\
&= \frac{1}{1-\bar{p}} \, .
\end{align}
By changing the mean membrane potential (e.g., by adjusting the leak potential or adding an external (bias) current), the probability of consecutive spikes $\bar{p}$ can be directly adjusted and hence, also the average length of bursts. Since these bursts are fixed structures with interspike intervals equal to the refractory period, they translate into side-peaks at multiples of the refractory period in the spike train ACF, as we demonstrate below. 

The ACF of the spike train $S$ is given by
\begin{equation}
\corr(S, S, \Delta)= \frac{\braket{S_t S_{t+\Delta}} - \braket{S}^2}{\mathrm{Var}(S)} \label{eq:SI_SCC} \,,
\end{equation} 
where the first term of the numerator is $\braket{S_t S_{t+\Delta}} = p(\mathrm{spike\ at\ }t+\Delta, \mathrm{spike\ at\ }t)$ (notation as in \cref{eq:SI_FMP,eq:SI_FMP2}). This term can be expressed as
\begin{align}
&p(\mathrm{spike\ at\ }t+\Delta, \mathrm{spike\ at\ }t) \nonumber \\ 
&= p(\mathrm{spike\ at\ }t+\Delta\ |\ \mathrm{spike\ at\ }t) \cdot p(\mathrm{spike\ at\ }t) \,, \label{eq:SI_CC}\\
&= p(\mathrm{spike\ at\ }t+\Delta\ |\ \mathrm{spike\ at\ }t) \cdot \braket{S} \,,
\end{align}
where we assumed that the first spike starts the burst at a random time $t$.
Therefore, in order to calculate the ACF, we have to calculate the probability that a spike occurs at time $t+\Delta$ given that the neuron spikes at time $t$.
This has to include every possible combination of spikes during this interval.
In the following, we argue that at multiples of the refractory period, the main contribution to the ACF comes from bursts.

\begin{itemize}
    \item
First, for $\Delta < \tauref$, the term $p(\mathrm{spike\ at\ }t+\Delta\ |\ \mathrm{spike\ at\ }t)$ in \cref{eq:SI_CC} vanishes since the neuron is refractory and cannot spike during this interval. 
Thus, the ACF becomes negative as only the term $-\frac{\braket{S}^2}{\mathrm{Var}(S)}$ in \cref{eq:SI_SCC} remains, where both numerator and denominator are positive. 
    \item
For $\Delta = \tauref$, a spike can only occur when the neuron bursts with probability $p_1 = \int_\vartheta^\infty \mathrm{d}u_{t+\tauref} f(u_{t+\tauref},\ \tauref\ |\ u_t = \vartheta)$, where we assumed for simplicity that the first spike starts the burst spontaneously. 
    \item
Since for $\tauref < \Delta < 2\tauref$, the neuron did not burst with probability $1-p_1$, it is possible to find a spike in this interval, leading again to negative, but diminished, values in the ACF.
    \item
For $\Delta = 2\tauref$, we now have two ways to observe spikes at $t$ and $t+2\tauref$: (i) The spikes are part of a burst of length 2 or (ii) there was no intermediate spike and the spikes have an ISI of $2\tauref$. Since for larger rates, having large ISIs that are exact multiples of $\tauref$ is unlikely, we can neglect the contribution of (ii).
    \item
If we go further to $\Delta = n\tauref$, we get even more additional terms including bursts of length $<n$. However, these terms can again be neglected as, compared to having a burst of length $n$, it is rather unlikely to get a burst pattern with missing intermediate spikes, i.e., having partial bursts which are a multiple of $\tauref$ apart.
    \item
Finally, for $\Delta \rightarrow \infty$, we have $\braket{S_t S_{t+\Delta}} - \braket{S}^2 = \braket{S_t}\braket{S_{t+\Delta}} - \braket{S}^2 = \braket{S}\braket{S} - \braket{S}^2 = 0$ and the ACF (\cref{eq:SI_SCC}) vanishes. 
\end{itemize}

Consequently, we can approximate the ACF at multiples of the refractory period by calculating the probability of finding a burst of $n$ spikes (\cref{eq:SIburst}):
\begin{align}
\corr(S, S, n\tauref) \approx \sum_{k=1}^\infty P_{k+1} \delta\big([n-k]\tauref\big) \,,
\end{align}
and for the special case of $\tausyn \ll \tauref$
\begin{align}
\corr(S, S, n\tauref) &\approx \sum_{k=1}^\infty \bar{p}^k \delta\big([n-k]\tauref\big) \\
&= \sum_{k=1}^\infty e^{k \ln{\bar{p}}} \delta\big([n-k]\tauref\big) \label{eq:SI_autocorr}\,.
\end{align}
Hence, since increasing the mean rate (or bias) of the neuron leads to an increase in $\bar{p}$ and thus to a reduced decay constant $\ln{\bar{p}}$, more significant side-peaks emerge. 

For $\tausyn \approx \tauref$, the effective membrane distribution is not yet stationary and therefore, this approximation does not hold.
To arrive at the exact solution, one would have to repeat the above calculation for all possible spike time combinations, leading to a recursive integral \cite{gerstner2002spiking}.
Furthermore, one would also need to take into account the situation where the first spike is itself part of a burst, i.e., is not the first spike in the burst.
To circumvent a more tedious calculation, we use an approximation which is in between the two cases $\tausyn \ll \tauref$ and $\tausyn \approx \tauref$: we use $\bar{p} = \int_\vartheta^\infty \mathrm{d}u f(u, \tauref\ |\ \vartheta)$, which provides a reasonable approximation for short bursts.

\subsection{Cross-correlation of free membrane potentials receiving correlated input}

Similarly to the ACF of the membrane potential, one can calculate the crosscorrelation function of the FMPs of two neurons receiving correlated noise input.
First, the membrane potentials are given by
\begin{align}
u_1 &= u_0^1 + S_\mathrm{tot, 1} * \kappa\,,\\
u_2 &= u_0^2 + S_\mathrm{tot, 2} * \kappa\,.
\end{align}
The covariance function can be written as
\begin{subequations}
\begin{align}
&\braket{\bar{u}_1(t)\bar{u}_2(t+\Delta)} \nonumber \\ 
&= \braket{u_1(t)u_2(t+\Delta)} - \braket{u_1(t)}\braket{u_2(t)}\\
&= \braket{\big(S_\mathrm{tot,1}*\kappa\big)(t)\big(S_\mathrm{tot,2}*\kappa\big)(t+\Delta)} \nonumber \\ 
&\ \ - \braket{\big(S_\mathrm{tot,1}*\kappa\big)(t)}\braket{\big(S_\mathrm{tot,2}*\kappa\big)(t)}\\
&= ... \nonumber \\
&= \big( \lim\limits_{T \rightarrow \infty} \frac{1}{T} \braket{\bar{S}_\mathrm{tot, 1}(t)\bar{S}_\mathrm{tot, 2}(t+\Delta')}_T \nonumber \\
&\ \ * \braket{\kappa(t)\kappa(t+\Delta')}_\infty\big)(\Delta)\,,
\end{align}
\end{subequations}
with $\bar{u} = u - \braket{u}$, from which we obtain the crosscorrelation function by normalizing with the product of standard deviations of $u_1$ and $u_2$ (for notation, see \cref{eq:SI_lim}). 
The term containing the input correlation coefficient is $\braket{\bar{S}_\mathrm{tot, 1}(t)\bar{S}_\mathrm{tot, 2}(t+\Delta')}$.
Plugging in the spike trains, we get four crosscorrelation terms
\begin{align}
&\braket{\bar{S}_\mathrm{tot, 1}(t)\bar{S}_\mathrm{tot, 2}(t+\Delta')} \nonumber \\
&= \sum_{l,m \in \{\mathrm{e}, \mathrm{i}\}} \Lambda_{l\mathrm{,1}}\Lambda_{m\mathrm{,2}} \braket{\bar{S}_{l\mathrm{,1}}(t)\bar{S}_{m\mathrm{,2}}(t+\Delta')}\,.
\end{align}
Since excitatory as well as inhibitory noise inputs are randomly drawn from the same pool of neurons, we can assume that $\braket{\bar{S}_{l\mathrm{,1}}(t)\bar{S}_{m\mathrm{,2}}(t+\Delta')}$ is approximately equal for all combinations of synapse types when averaging over enough inputs, regardless of the underlying correlation structure/distribution of the noise pool.
The first term, however, depends on the synapse types since the $\Lambda$-terms (\cref{eq:SI_lambda}) contain the distance between reversal potentials and mean FMP:
\begin{align}
&\braket{\bar{u}_1(t)\bar{u}_2(t+\Delta)}\nonumber \\ 
&= \zeta_1 \zeta_2 \sum_{l,m \in \{\mathrm{e}, \mathrm{i}\}} w_l w_m \big(E_l^\mathrm{rev}- \mu_1\big)\big(E_m^\mathrm{rev}- \mu_2\big) \nonumber \\ 
&\cdot \bigg[\braket{\bar{S}_{l\mathrm{,1}}(t)\bar{S}_{m\mathrm{,2}}(t+\Delta')} * \braket{\kappa(t)\kappa(t+\Delta')}\bigg](\Delta) \label{eq:SI_crosscorr} \,, 
\end{align}
with constants $\zeta_i = \frac{\tausyn}{\braket{\gtot_i}}\big(\tausyn-\braket{\taueff^i}\big)$.
The cross-correlation vanishes when, after summing over many inputs, the following identities hold:
\begin{subequations}
\begin{align}
\braket{\Lambda_{\mathrm{e,1}}\Lambda_{\mathrm{e,2}}}_\mathrm{inputs} &= -\braket{\Lambda_{\mathrm{e,1}}\Lambda_{\mathrm{i,2}}}_\mathrm{inputs}\,,\\
\braket{\Lambda_{\mathrm{i,1}}\Lambda_{\mathrm{i,2}}}_\mathrm{inputs} &= -\braket{\Lambda_{\mathrm{i,1}}\Lambda_{\mathrm{e,2}}}_\mathrm{inputs}\,,
\end{align}
\end{subequations}
where $\braket{(\cdot)}$ is an average over all inputs, i.e., all neurons that provide noise.

While not relevant for our simulations, it is worth noting that the excitatory and inhibitory weights with which each neuron contributes its spike trains can be randomly drawn from non-identical distributions.
By enforcing the following correlation between the noise weights of both neurons, one can introduce a skew into the weight distribution which compensates for the differing distance to the reversal potentials:
\begin{align}
&(E_\mathrm{rev}^\mathrm{e,1} - \mu_1)(E_\mathrm{rev}^\mathrm{e,2} - \mu_2)\braket{w_\mathrm{e}^1w_\mathrm{e}^2}_\mathrm{inputs} \nonumber \\ 
&= -(E_\mathrm{rev}^\mathrm{e,1} - \mu_1)(E_\mathrm{rev}^\mathrm{i,2} - \mu_2)\braket{w_\mathrm{e}^1w_\mathrm{i}^2}_\mathrm{inputs} \label{eq:eq2}
\end{align}

A simple procedure to accomplish this is the following: First, we draw the absolute weights $w^1$ and $w^2$ from an arbitrary distribution and assign synapse types randomly with probabilities $p_\mathrm{e/i}$ afterwards.
If $w^2$ is excitatory, we multiply $w^1$ by $\frac{|E_\mathrm{rev}^{i,2}-\mu_2|}{p_e |E_\mathrm{rev}^{i,2}-\mu_2| + p_i |E_\mathrm{rev}^{e,2}-\mu_2|}$, otherwise by $\frac{|E_\mathrm{rev}^{e,2}-\mu_2|}{p_e |E_\mathrm{rev}^{i,2}-\mu_2| + p_i |E_\mathrm{rev}^{e,2}-\mu_2|}$.
This way, $\braket{w^1}$ remains unchanged and the resulting weights suffice \cref{eq:eq2}.

\subsection{State space switch from \{0,1\} to \{-1,1\}}

To switch from the state space $\bs z \in \{0,1\}$ to $\bs z' \in \{-1, 1\}$ while conserving the state probabilities (i.e., $p(\bs z) = p(\bs z')$) one has to adequately transform the distribution parameters $\bf W$ and $\bf b$. 
Since the distributions are of the form $p(\bs z) = \exp\left(\bs z^T \textbf{W} \bs z + \bs z^T \textbf{b}\right)$, this is equivalent to requiring that the energy $E(\bs z) = \bs z^T \textbf{W} \bs z + \bs z^T \textbf{b}$ of each state remains, up to a constant, unchanged.

First, we can write the energy of a state $\bs z'$ and use the transformation $\bs z' = 2\bs z - 1$ to get
\begin{subequations}
\begin{align}
E(\bs z') &= \frac{1}{2} \sum_{i,j} \bs z'_i \textbf{W}'_{ij} \bs z'_j + \sum_i \bs z'_i \textbf{b}'_i \\
&= \frac{1}{2} \bigg(4 \sum_{i,j} \bs z_i \textbf{W}'_{ij} \bs z_j - 2 \sum_{i,j} \bs z_i \textbf{W}'_{ij} - 2 \sum_{i,j} \textbf{W}'_{ij} \bs z_j \nonumber \\
&\ \ + \sum_{i,j} \textbf{W}'_{ij} \bigg) - \sum_i \textbf{b}'_i + 2 \sum_i \bs z_i \textbf{b}'_i \\
&= \frac{1}{2} \sum_{i,j} \bs z_i 4 \textbf{W}'_{ij} \bs z_j + \sum_i \bs z_i \big(2 \textbf{b}'_i \nonumber \\
&\ \ - 2 \sum_j \textbf{W}'_{ij}\big) + C \, ,
\end{align}
\end{subequations}
where $C$ is a constant $C = \frac{1}{2} \sum_{i,j} \textbf{W}'_{ij} - \sum_i \textbf{b}'_i$ and we used the fact that $\textbf{W}'_{ij}$ is symmetric.
Since constant terms in the energy leave the probability distribution invariant, we can simply compare $E(\bs z')$ and $E(\bs z)$
\begin{equation}
E(\bs z) = \frac{1}{2} \sum_{i,j} \bs z_i^T \textbf{W}_{ij} \bs z_j + \sum_i \bs z_i^T \textbf{b}_i\,, \\
\end{equation}
and extract the correct parameter transformation:
\begin{align}
\textbf{W}_{ij} &= 4 \textbf{W}'_{ij}\,, \\
\textbf{b}_i &= 2 \textbf{b}'_i - 2 \sum_j \textbf{W}'_{ij}\,.
\end{align}
From this, we can also calculate the inverse transformation rule for $\bs z = \frac{1}{2}(\bs z' + 1)$:
\begin{align}
\textbf{W}'_{ij} &= \frac{1}{4} \textbf{W}_{ij}\,, \label{SI:eq1}\\
\textbf{b}'_i &= \frac{1}{2} \textbf{b}_i + \frac{1}{4} \sum_j \textbf{W}_{ij} \label{SI:eq2}\,.
\end{align}

\subsection{Translation from Boltzmann to neurosynaptic parameters}

As discussed in the methods section, following \cite{petrovici2016stochastic}, the activation function of LIF neurons in the HCS is approximately logistic and can be written as
\begin{align}
p(\bs z_k = 1\ |\ \bs z_{/k}) &= \varphi(\mu) \nonumber \\ 
&= \left(1+\exp{(-(\mu-u_0)/\alpha})\right)^{-1} \label{eq:SI_motivate_transl}\,, 
\end{align}
where $\bs z_{/k}$ is the state vector of all other neurons except the k'th one and $\mu$ the mean membrane potential (\cref{eq:SI_mu}).
$u_0$ and $\alpha$ are the inflection point and the inverse slope, respectively.
Furthermore, the conditional probability $p(\bs z_k = 1\ |\ \bs z_{/k})$ of a Boltzmann distribution over binary random variables $\bs z_k$, i.e., $p(\bs z) \propto \exp\left( \frac{1}{2} \bs z^T \bs W \bs z + \bs z^T \bs b \right)$, is given by 
\begin{align}
&p(\bs z_k = 1\ |\ \bs z_{/k}) \nonumber \\ 
&= \left(1+\exp{(-\sum_j W_{kj}\bs z_j - b_k })\right)^{-1} \,,
\end{align}
with symmetric weight matrix $\bs W$, $W_{ii} = 0$ $\forall i$, and biases $\bs b$. These equations allow a translation from the parameters of a Boltzmann distribution $(b_i,\ W_{ij})$ to parameters of LIF neurons and their synapses $(\El,\ w_{ij})$, such that the state dynamic of the network approximates sampling from the target Boltzmann distribution.

First, the biases $\bs b$ can be mapped to leak potentials $\El$ (or external currents) by requiring that, for $\bs W = 0$ (that is, no synaptic input from other neurons), the activity of each neuron equals the conditional probability of the target Boltzmann distribution
\begin{equation}
\left(1+\exp{(-(\mu-u_0)/\alpha})\right)^{-1} \overset{!}{=} \left(1+\exp{(- b_k })\right)^{-1} \,,
\end{equation}
leading to the translation rule
\begin{equation}
 \bs \El = \frac{\tau_\mathrm{m}}{\taueff} (\alpha \bs b + u_0) - \sum_{x\in \{\mathrm{e},\mathrm{i}\}} \frac{\braket{\gsyn_x}}{\gl} \Erev_x \label{eq:SI_biastrans} \,.
\end{equation}

To map Boltzmann weights $W_{ij}$ to synaptic weights $w_{ij}$, we first have to rescale the $W_{ij}$, as done for the biases in \cref{eq:SI_biastrans}.
However, leaky integrator neurons have non-rectangular PSPs, so their interaction strength is modulated over time.
This is different from the interaction in Boltzmann machines, where the PSP shape is rectangular (Glauber dynamics).
Nevertheless, we can derive a heuristic translation rule by requiring that the mean interaction during the refractory period of the presynaptic neuron is the same in both cases, i.e.,
\begin{subequations}
\begin{align}
\int_0^\tauref \mathrm{d}t\ PSP(t) &\overset{!}{=} \int_0^\tauref \mathrm{d}t\ \alpha W_{ij} \\
&= \alpha W_{ij} \tauref \,,
\end{align}
\end{subequations}
where $PSP(t)$ is given by \cref{eq:SI_PSP}.
From this, we get the translation rule for synaptic weights:
\begin{equation}
 w_{kj} = \frac{\alpha W_{kj} \Cm \frac{\tauref}{\tausyn} \left(1-\frac{\tausyn}{\taueff}\right) \left(\Erev_{kj} - \mu\right)^{-1}}{ \left[\tausyn \left(e^{-\frac{\tauref}{\tausyn}} - 1 \right) - \taueff \left( e^{- \frac{\tauref}{\taueff}} - 1 \right) \right]} \label{eq:SI_w_rule}\, .
\end{equation}

%% file: tex_files/acknowledgments.tex
We thank Luziwei Leng, Nico G{\"u}rtler and Johannes Bill for valuable discussions. 
We further thank Eric M{\"u}ller and Christian Mauch for maintenance of the computing cluster we used for simulations and Luziwei Leng for providing code implementing the CAST algorithm.
This work has received funding from the European Union 7th Framework Programme under grant agreement 604102 (HBP), the Horizon 2020 Framework Programme under grant agreement 720270, 785907 (HBP) and the Manfred St{\"a}rk Foundation.

%% file: supplement/sim_details.tex
\subsection{Experiment details}

All simulations were done with PyNN 0.8 and NEST 2.4.2. 
The LIF model was integrated with a time step of $\mathrm{dt} = \SI{0.1}{ms}$. 
Since the SSN is a time-continuous system, we could, in principle, retrieve a sample at every integration step. 
However, as individual neurons only change their state on the time scale of refractory periods, and hence new states emerge on a similar time scale, we read out the states in intervals of $\frac{\tauref}{2}$.
If not stated otherwise, we used $U_\mathrm{SE} = 1.0$ and $\taurec = \SI{10}{ms}$ as short term plasticity parameters for connections within each SSN to ensure postsynaptic potentials with equal height, as discussed in \cite{petrovici2016form}.
For background connections, i.e., Poisson input or background input coming from other SSNs in an ensemble, we used static synapses ($U_\mathrm{SE} = 1.0$ and $\taurec \rightarrow 0$) instead to facilitate the mathematical analysis.
For the interconnections of an ensemble, we expect that short-term depression will not alter the performance of individual SSNs in a drastic way, as the effect will be rather small on average if most neurons are far away from tonic bursting.
Thus, to allow a clear comparability to SSNs receiving Poisson input, we chose to neglect short-term depression for ensemble interconnections.
The neuron parameters used throughout all simulations are listed in Table S1.

\subsubsection{Details to Figure 2B,C,D of main text}

The parameters for the target distributions of all networks were randomly drawn from beta distributions, i.e., $W \sim 2\cdot(\mathrm{beta}(0.5, 0.5) - 0.5)$ and $b \sim 1.2\cdot(\mathrm{beta}(0.5, 0.5) - 0.5)$.
The bias of each background-providing neuron was adjusted to yield the desired firing rate $p \in \{0.1, 0.6, 0.9\}$ in case of no synaptic input ($W=0$) $b = \mathrm{log}\left(\frac{p}{1-p}\right)$.
For the different activity cases, we used $N_{0.1} = 260$, $N_{0.6} = 50$ and $N_{0.9} = 34$ networks as background input for each neuron to reach the desired background noise frequency.
The Poisson frequency of the noise-providing networks was set to \SI{3000}{Hz}.
Activation functions were recorded by providing every neuron with background noise for \SI{5e5}{ms} and varying its leak potential.
For the autocorrelation functions, we first merged all individual noise spike trains and binned the resulting spike train with a bin size of \SI{0.5}{ms} before calculating the autocorrelation function.

\subsubsection{Details to Figure 2F,G of main text}

Background input was generated from a pool of pairwise connected neurons (i.e., small subnetworks with two neurons each) with strong positive or negative weights to yield highly positively or negatively correlated spike trains.
Each pair of neurons in the main network (i.e, the network receiving no Poisson input) received the spikes of 80 such subnetworks as background input.
The weights of the noise-generating subnetworks were drawn from beta distributions $w^\mathrm{pre}_{ij} \sim 4\cdot(\mathrm{beta}(5.0, 0.5) - 0.5)$ (distribution strongly skewed to positive weights) or $w^\mathrm{pre}_{ij} \sim 4\cdot(\mathrm{beta}(0.5, 5.0) - 0.5)$ (skewed to negative weights).
The parameters of the main network were randomly generated as in \cref{fig:fig2}D.
The absolute values of the weights $W_\mathrm{noise}$ projecting from the noise-generating subnetworks to the main network were randomly generated from a (Gaussian-like) beta distribution $W_\mathrm{noise} \sim (\mathrm{beta}(4.0, 4.0)-0.5)\cdot2\cdot0.001 + \SI{0.001}{\micro\siemens}$.
The synapse type of each weight $W_\mathrm{noise}$ was determined randomly with equal probability.
Furthermore, inhibitory synapses were scaled by a factor of $1.35$ such that inhibitory and excitatory weights have the same mean value as for the simulations with Poisson noise (see Table S1).
For the three traces shown, the absolute value of each synaptic weight was drawn independently.
Synapse types were either drawn according to a pattern (\cref{fig:fig2}F, left and middle) or independently (\cref{fig:fig2}F, right).

\subsubsection{Details to Figure 3A,B,C of main text and Figure S2 in the Supporting information}

For every network, the target distribution parameters were again drawn from a beta distribution as in \cref{fig:fig2}.
The connectivity of the noise connections was set to $\epsilon = 0.05$, i.e., each neuron received the spikes of $5$\% of the remaining subnetworks' neurons as stochastic background input, for the ensemble of 400 3-neuron networks (no training, Figure S2 in the Supporting information) and to $\epsilon = 0.10$ for the ensemble of 100 6-neuron networks (training, main text).

The training was done for subnetworks with 6 neurons each, where every subnetwork was initialized with different weights and biases than the target parameters, also generated randomly. 
As an initial guess for the neurons' activation functions, we used the activation function of a neuron receiving \SI{2000}{Hz} excitatory and inhibitory Poisson input, leading to a slope of $\alpha = \SI{1.47}{mV}$ and a mid-point at \SI{-52.97}{mV}.
These parameters were subsequently used to translate the weight and bias updates given by the Hebbian wake-sleep algorithm (see main text, Eqn. 8 and 9) to updates of synaptic weights and leak potentials.
The subnetworks were all trained simultaneously with contrastive divergence, where the model term was approximated by sampling for \SI{1e5}{ms}.
The training was done for 2000 steps and with a learning rate of $\frac{400}{t+2000}$.
As a reference, 50 subnetworks receiving only Poisson noise (\SI{2000}{Hz}) were also trained in the same way for 2000 steps.

Self-activation of the network can be observed when a large enough fraction of neurons have a suprathreshold rest potential, in our case around 30\%.

\subsubsection{Details to Figure 3D,E of main text}

The emulated ensemble consists of 15 4-neuron networks which were randomly initialized on two HICANN chips (HICANN 367 and 376 on Wafer 33 of the BrainScaleS system). 
The analog hardware parameters which determine the physical range of weights adjustable by the 4bit setting were set to $gmax=500$ and $gmax\_div=1$. 
Given the current state of development of the BrainScaleS system and its surrounding software, we limited the experiment to small ensembles in order to avoid potential communication bottlenecks.

Biases were implemented by assigning every SSN neuron a bias neuron, with its rest potential set above threshold to force continuous spiking.
While a more resource-efficient implementation of biases is possible, this implementation allowed an easier mapping of neuron - bias pairs on the neuromorphic hardware.
Bias strengths can then be adjusted by modifying the synaptic weights between SSN neurons and their allocated bias neurons.
The networks were trained with the contrastive divergence learning rule to sample from their respective target distributions.
Biases were randomly drawn from a normal distribution with $\mu = 0$ and $\sigma = 0.25$.
The weight matrices were randomly drawn from $W \propto 2\cdot(\mathrm{beta}(0.5, 0.5) - 0.5)$ and subsequently symmetrized by averaging with their respecetive transposes $0.5\cdot (W+W^\mathrm{T})$.

Since the refractory periods of hardware neurons vary, we further measured the refractory period of every neuron in the ensemble, which was later used to calculate the binary neuron states from spike raster plots.
Refractory periods were measured by setting biases to large enough values to drive neurons to their maximal firing rate. 
After running the experiment, the duration of the refractory period can be approximated by dividing the experiment time by the number of measured spikes. 

During the whole experiment, the ensemble did not receive external Poisson noise.
Instead, individual SSNs received spikes from 20\% of the remaining ensemble as background input, with noise connections having hardware weights of $\pm 4$ with the sign chosen randomly with equal probability.
Translation between theoretical and 4bit hardware parameters was done by clipping the values into the range $[-15, 15]$ and rounding to integer values.
Calculation of weight and bias updates was performed on a host computer.
The learning rate was set to $\eta = 1.0$ for all networks performing better than the median and twice this value for networks performing worse.
For every training step, the ensemble was recorded for \SI{1e5}{ms} biological time before applying a parameter update.

For the experiments with Poisson noise, every neuron received \SI{300}{Hz} external Poisson noise provided by the host computer. 

The neuron parameters used for all hardware experiments are listed in Table S2.

\subsubsection{Details to Figure 4 of main text and Figure S4-6 in the Supporting information}

To reduce the training time, we pretrained classical restricted Boltzmann machines on their respective datasets, followed by direct translation to spiking network parameters.
To obtain better generative models, we utilized the CAST algorithm \cite{salakhutdinov2010learning} which combines contrastive divergence with simulated tempering.
Each subnetwork was trained for 200000 steps with a minibatch size of 100, a learning rate of $\frac{20}{t+2000}$, an inverse temperature range $\beta \in [1., 0.6]$ with 20 equidistant intervals and an adaptive factor $\gamma_t = \frac{9}{1+t}$.
States between the fast and slow chain were exchanged every 50 samples.
To collect the background statistics of these subnetworks, we first simulated all networks with stochastic Poisson input.
To improve the Poisson-stimulated reference networks' mixing properties, we utilized short-term depression to allow an easier escape from local energy minima faster ($U_\mathrm{SE} = 0.01$, $\taurec = \SI{280}{ms}$, global weight rescale $\delta W = 0.014^{-1}$). 
For classification, the grayscale value of image pixels was translated to spiking probabilities of the visible units, which can be adjusted by setting the biases as $\mathrm{ln}\bigg(\frac{\mathrm{grey\ value}}{1-\mathrm{grey\ value}}\bigg)$.
Spike probabilities of 0 and 1 were mapped to biases of -50 and 50.
Furthermore, during classification, the connections projecting back from the hidden neurons to the visible neurons were silenced in order to prevent the hidden layer from influencing the clamped input.
For pattern rivalry, the non-occluded pixels were binarized.
In total, each SSN received background input from 20\% of the other networks' hidden neurons.
For the classification results, the experiment was repeated 10 times for different random seeds (leading to different connectivity matrices between the SSNs).
For training and testing, we used 400 and 200 images per class. 
In \cref{fig:fig4}D, consecutive images are \SI{400}{ms} apart.
In \cref{fig:fig4}E, for the clamped "B", consecutive images are 2s apart, for the "L" 1.5s.
	
The experiments with MNIST used an ensemble of five networks with 784 visible neurons and 500 hidden neurons each (\cref{fig:figS5} and Video S2 in the Supporting information), trained on $6\cdot10^3$ images per digit class (where we took the digits provided by the EMNIST set to have balanced classes).
Since the generative properties of larger SSNs depend heavily on the synaptic interaction, we also used short-term plasticity for the case without Poisson noise ($U_\mathrm{SE} = 0.01$, $\taurec = \SI{280}{ms}$, $\delta W = 0.01^{-1}$) to allow fluent mixing between digit classes \cite{leng2018spiking}.
For MNIST, each SSN received background input from 30\% of the other networks' hidden neurons.
Furthermore, the excitatory noise weight was set to $w^\mathrm{noise}_\mathrm{e} = \SI{0.0009}{\micro\siemens}$.

The network-generated images were obtained by averaging the firing activity of the visible neurons (for pattern rivalry $\pm \SI{90}{ms}$, for classification and dreaming of EMNIST $\pm \SI{80}{ms}$ and for MNIST $\pm \SI{140}{ms}$).
The time intervals were chosen to reduce the blur caused by mixing when plotting averaged spiking activity.\\

%% file: supplement/Table_S1.tex
\begin{table}[H]\caption{Neuron parameters used in simulations throughout the main text. The membrane time constant was chosen small such that smaller noise frequencies suffice to reach a high-conductance state, allowing us to use smaller ensembles and hence reduce simulation time.}\centerline{\begin{tabular}{ccc}
	\toprule
    $\Cm$       & 0.1 nF & membrane capacitance\\	
	$\tau_\mathrm{m}$    & 1.0 ms & membrane time constant \\
	$E_{\mathrm{l}}$ &-65.0 mV & leak potential\\
	$E^{\mathrm{rev}}_{\mathrm{e}}$ & 0.0 mV & exc. reversal potential  \\
	$E^{\mathrm{rev}}_{\mathrm{i}}$ & -90.0 mV & inh. reversal potential \\
	$\vartheta$ & -52.0 mV & threshold potential\\
	$\varrho$ & -53.0 mV & reset potential \\
	$\tau^{\mathrm{syn}}_{\mathrm{e}}$ & 10.0 ms & exc. synaptic time constant \\
	$\tau^{\mathrm{syn}}_{\mathrm{i}}$ & 10.0 ms & inh. synaptic time constant\\
	$\tau_{\mathrm{ref}}$ & 10.0 ms & refractory time\\
	$w^{\mathrm{noise}}_{\mathrm{e}}$ & 0.001 $\mu$S & exc. Poisson weights\\
	$w^{\mathrm{noise}}_{\mathrm{i}}$ & -0.00135 $\mu$S & inh. Poisson weights\\
	\bottomrule
\end{tabular}}
\end{table} 

%% file: supplement/Table_S2.tex
\begin{table}[H]\caption{Neuron parameters used for the implementations in an artificial neural substrate. 
Note that these are intended parameters and the realized ones can vary from neuron to neuron.}\centerline{\begin{tabular}{ccc}
	\toprule
     & ensemble neurons & bias neurons\\  
    \midrule
    $\Cm$       & 0.2 nF & \\	
	$\tau_\mathrm{m}$    & 7 ms &  \\ 
	$E_{\mathrm{l}}$ &-20.0 mV & 60.0 mV\\
	$E^{\mathrm{rev}}_{\mathrm{e}}$ & 60.0 mV &   \\
	$E^{\mathrm{rev}}_{\mathrm{i}}$ & -100.0 mV &  \\
	$\vartheta$ & -20.0 mV & \\
	$\varrho$ & -35.0 mV & -30.0 mV \\
	$\tau^{\mathrm{syn}}_{\mathrm{e}}$ & 8.0 ms & 5.0 ms \\
	$\tau^{\mathrm{syn}}_{\mathrm{i}}$ & 8.0 ms & 5.0 ms\\
	$\tau_{\mathrm{ref}}$ & 4.0 ms & 1.5 ms\\
	\bottomrule
\end{tabular}}
\end{table} 

%% file: supplement/movies.tex
\subsection{Video captions}

\subsubsection{Video S1} 
Sampled distribution over time from an autonomous ensemble (no explicit noise) of 15 4-neuron networks on an artificial neural substrate (the BrainScaleS system). (top) Median $D_\mathrm{KL}$ of the ensemble (red) and the individual networks (red, transparent) as a function of time after training. The median $D_\mathrm{KL}$ pre-training is shown in black. (Bottom) Comparison between target distribution (blue) and sampled distribution (red) for all networks. Most networks are able to approximate their target distribution well (e.g., the networks at position (1,1), (4,1) and (4,2) with (\textit{row}, \textit{column})) or at least approximate the general shape of the target distribution (e.g., (0,1) and (0,2)).  The networks at position (2,2) and (3,0) show strong deviations from their respective target distributions due to single neuron deficiencies. Because of the speed-up of the BrainScaleS system, it only takes 100ms to emulate $10^6$ms of biological time.\\

\subsubsection{Video S2} 
Video of the data shown in \cref{fig:figS5} of the Supporting information. An ensemble of five hierarchical networks with 784-500-10 (visible-hidden-label) neurons trained on the MNIST handwritten dataset generating digits without explicit noise sources is shown (in simulation). Every network in the ensemble receives the spiking activity from hidden neurons of the other networks as stochastic input only. (top) Activity of the hidden layer of each network. (middle) Class label currently predicted by the label neurons. (Bottom) Activity of the visible layer. To generate grayscale images from spikes, we averaged the spiking activity of each neuron over a window of size $\pm 90$ms.

\subsubsection{Video S3} 
Video of the data shown in \cref{fig:figS7} of the Supporting information. A single hierarchical network with 784-200 (visible-hidden) neurons generating samples of the MNIST handwritten digits dataset without explicit noise sources is shown (in simulation). To initialize the network, we trained a Boltzmann machine and translated the weights and biases to neurosynaptic parameters to reduce simulation time. (top) Illustration of the used network architecture. Lateral (non-plastic) connections in each layer were utilized as a noise source (red), with an interconnectivity of $\epsilon = 0.2$. (bottom) Averaged activity (average window $\pm 90$ms) of the visible layer (left) after initializing the network, (middle) after further training the network and (right) for the case of explicit Poisson noise instead of lateral interconnections. After initialization, the network is able to generate recognizable images but does not mix well between different digit classes. Further training the network on images of the MNIST training set improves both image quality and mixing, rivaling the quality of the reference setup with explicit Poisson noise. During the second training phase, neurosynaptic parameters are adjusted such that every neuron is able to perform its task with the available background activity it receives.

%% file: supplement/SI_figures.tex
\vspace*{\fill}
\begin{figure*}[h]
    \centering
    \includegraphics[width=\textwidth]{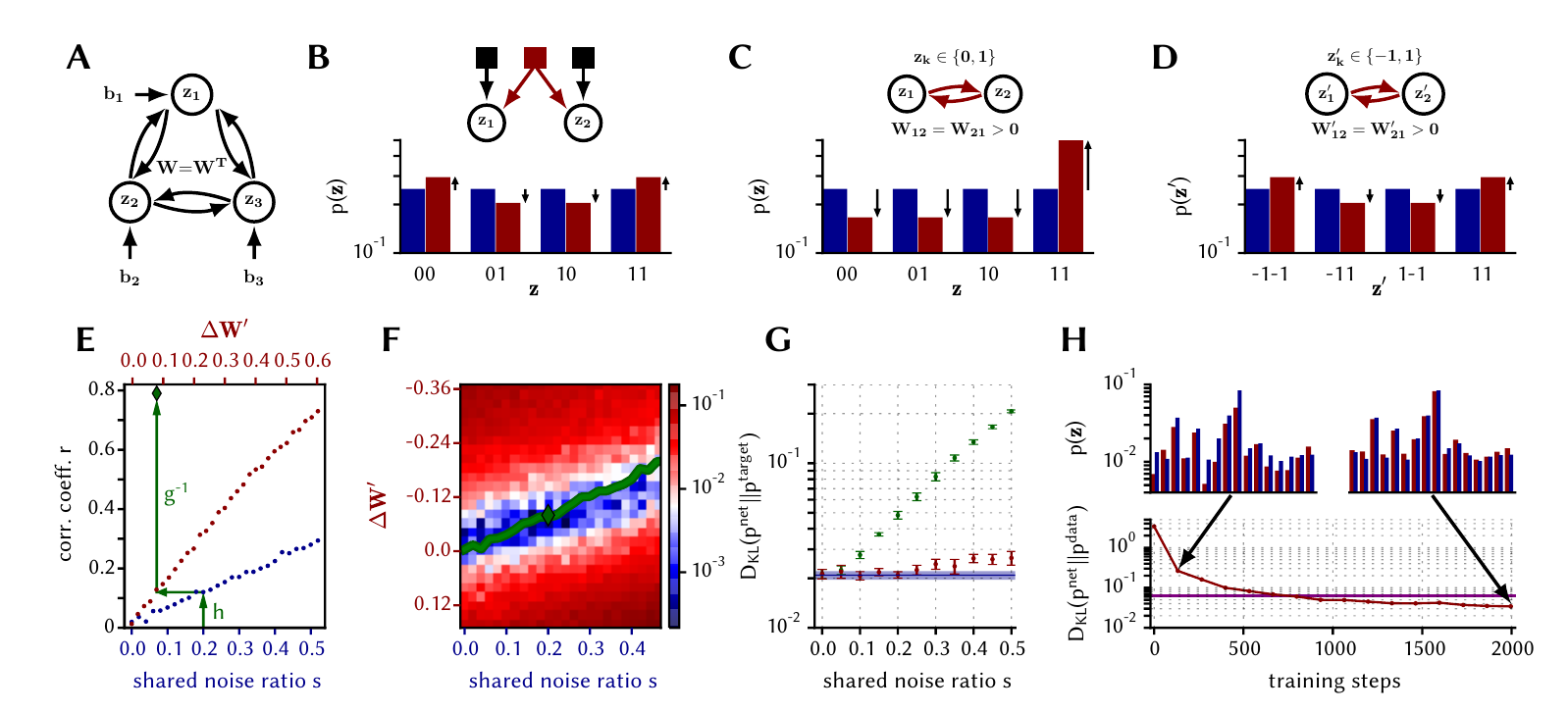}
    \caption{
    	Compensation of input correlations by adjustment of weights and biases in an SSN \cite{cns2017,bytschok2017spike}.
    	For simplicity, this is illustrated here for the case of shared input correlations, but the results hold for all types of statically correlated inputs.
        \textbf{(A)} Exemplary architecture of a network with 3 neurons that samples from a Boltzmann distribution with parameters $\mathbf{W}$ and $\mathbf{b}$.
            In order to achieve the required stochastic regime, each neuron receives external noise in the form of Poisson spike trains (not shown).
        \textbf{(B)-(D)} Exemplary sampled distributions for a network of two neurons.
            The ``default'' case is the one where all weights and biases are set to zero (uniform distribution, blue bars).
        \textbf{(B)} Shared noise sources have a correlating effect, shifting probability mass into the (1,1) and (0,0) states (red bars).
        \textbf{(C)} In the $\{0,1\}^2$ space, increased weights introduce a (positive) shift of probability mass from all other states towards the (1,1) state (red bars), which is markedly different from the effect of correlated noise.
        \textbf{(D)} In the $\{-1,1\}^2$ space, increased weights have the same effect as correlated noise (red bars).
        \textbf{(E)} Dependence of the correlation coefficient $r$ between the states of two neurons on the change in synaptic weight $\Delta W'$ (red) and the shared noise ratio $s$ (blue).
            These define bijective functions $g$ and $h$ that can be used to compute the weight change ($\Delta W' = f(s), \quad \text{with} \; f := g^{-1} \circ h$) needed to compensate the effect of correlated noise in the $\{-1,1\}^N$ space.
        \textbf{(F)} Study of the optimal compensation rule in a network with two neurons.
            For simplicity, the ordinate represents weight changes for a network with states in the $\{-1,1\}^2$ space, which are then translated to corresponding parameters ($\mathbf{W}, \mathbf{b}$) for the $\{0,1\}^2$ state space.
            The colormap shows the difference between the sampled and the target distribution measured by the Kullback-Leibler divergence $\DKL{\pnet}{\ptarget}$.
            The mapping provided by the compensation rule $f$ (see (E)) is depicted by the green curve.
            Note that the compensation rule $\Delta W' = f(s)$ provides a nearly optimal parameter translation.
            Remaining deviations are due to differences between LIF and Glauber dynamics.
        \textbf{(G)} Compensation of noise correlations in an SSN with ten neurons.
            The results are depicted for a set of ten randomly drawn Boltzmann distributions over $\bs z \in \{0,1\}^{10}$ (error bars).
            For a set of randomly chosen Boltzmann distributions, a ten-neuron network performs sampling in the presence of pairwise-shared noise ratios $s$ (x-axis).
            The blue line marks the sampling performance without noise-induced correlations ($s=0$).
            For an increasing shared noise ratio, uncompensated noise (green) induces a significant increase in sampling error.
            After compensation, the sampling performance is nearly completely restored.
            As before, remaining deviations are due to differences between LIF and Glauber dynamics.
        \textbf{(H)} An LIF-based ten-neuron network with shared noise sources ($s=0.3$ for each neuron pair) is trained with data samples generated from a target Boltzmann distribution (blue bars).
            During training, the sampled distribution becomes an increasingly better approximation of the target distribution (red line).
            For comparison, we also show the distribution sampled by an SSN with parameters translated directly from the Boltzmann parameters (purple).
            The trained network is able to improve upon this result because learning implicitly compensates for the abovementioned differences between LIF and Glauber dynamics.
            } \label{fig:figS1}
\end{figure*}
\vspace*{\fill}
\clearpage
\vspace*{\fill}
\begin{figure*}[h]
    \centering
    \includegraphics[width=\textwidth]{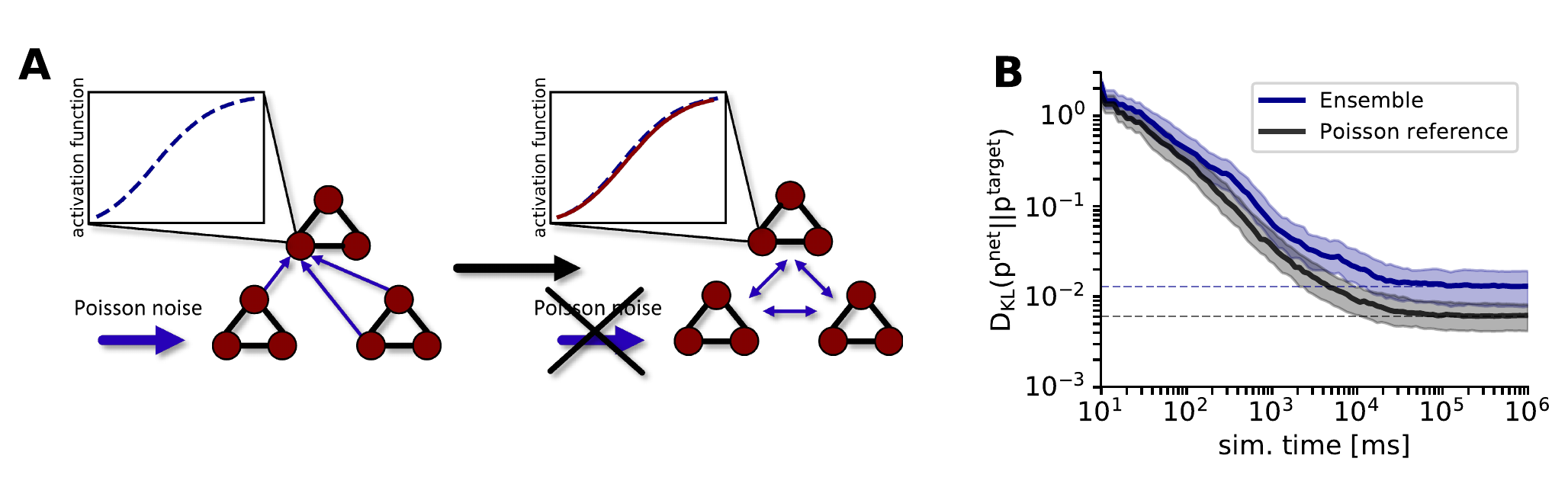}
    \caption{
        Translating the parameters of target distributions to neurosynaptic parameters. \textbf{(A)} A straightforward way to set up the parameters of each network ($w_{ij}$ and $E_l$) is to use the parameter translation as described in the main text, i.e., use the corresponding
        activation function of each neuron to correctly account for the background noise statistics.
        This is demonstrated here for the case of (left) 399 networks (only two shown) receiving Poisson noise and one network only receiving ensemble input and (right) all networks only receiving ensemble input.
        In both cases, the resulting activation function is the same and we can indeed use it to translate the parameters of the target distribution to neurosynaptic parameters.
        \textbf{(B)} Using the corresponding activation functions to set up the ensemble (but no training), each network in the ensemble is indeed able to accurately sample from its target distribution without explicit noise, as expected from our considerations in (A) and the main text. 
        This is shown here (in software simulations) for an ensemble of 400 3-neuron SSNs with an interconnection probability of $0.05$, reaching a median $\DKLsolo$ of $12.8^{+6.4}_{-5.0}\times 10^{-3}$ (blue), which is close to the ideal result with Poisson noise of $6.2^{+2.0}_{-2.0}\times 10^{-3}$ (black, errors given as the first and third quartile).
} \label{fig:figS2}
\end{figure*}
\vspace*{\fill}
\begin{figure*}[h]
    \centering
    \includegraphics[width=\textwidth]{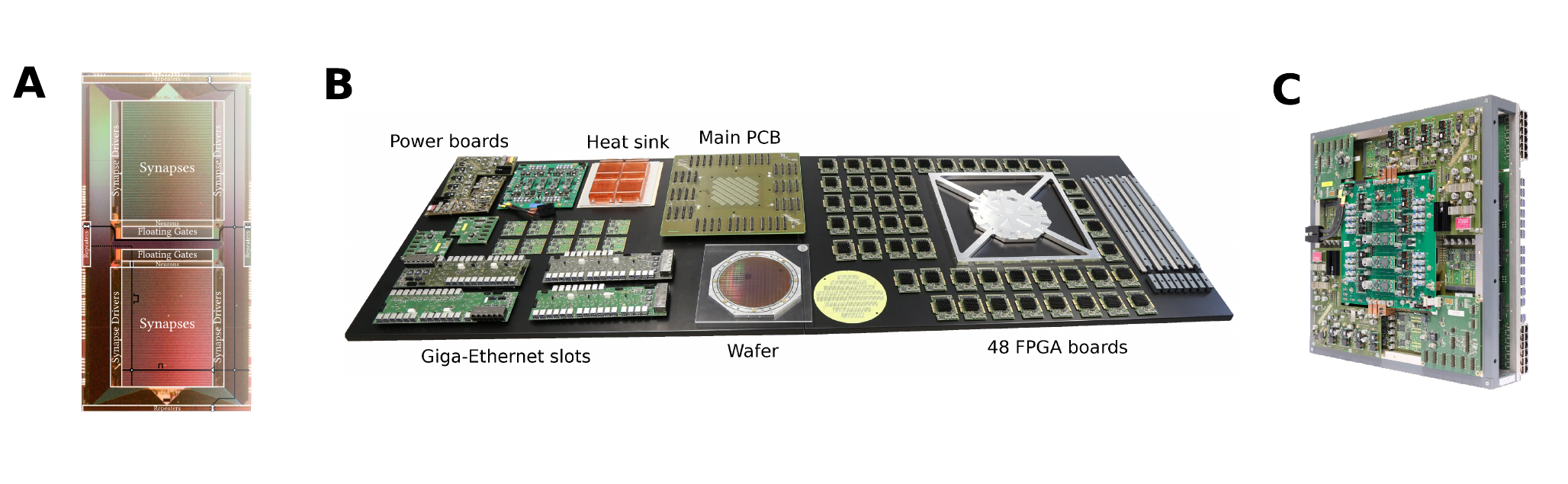}
    \caption{
        The BrainScales neuromorphic system. \textbf{(A)} A single HICANN chip (High Input Count Analog Neural Network), the elemental building block of the BrainScaleS wafer.
                    The HICANN consists of two symmetric halves and harbors analog implementations of adaptive exponential integrate-and-fire (AdEx) neurons and conductance-based synapses in 180nm CMOS technology.
                    Floating gates next to the neuron circuits are used to store neuron parameters. 
                    Spikes are routed digitally through horizontal and vertical buses (not shown) and translated into postsynaptic conductances in the synapse array.
                    Unlike in simulations on general-purpose CPUs, here neurons and synapses are physically implemented, with no numeric computations being performed to calculate network dynamics.
                    A single wafer consists of 384 HICANN chips.
        \textbf{(B)} Individual components of the BrainScaleS system, including both wafer and support structure.
        For instance, FPGA boards provide an I/O interface for wafer configuration and spike data and Giga-Ethernet slots provide a connection between FPGAs and the control cluster from which users conduct their experiments via Python scripts using the PyNN API.
        \textbf{(C)} Completely assembled wafer of the BrainScaleS neuromorphic system.
} \label{fig:figS3}
\end{figure*}
\vspace*{\fill}
\clearpage
\vspace*{\fill}
\begin{figure*}[h]
    \centering
    \includegraphics[width=\textwidth]{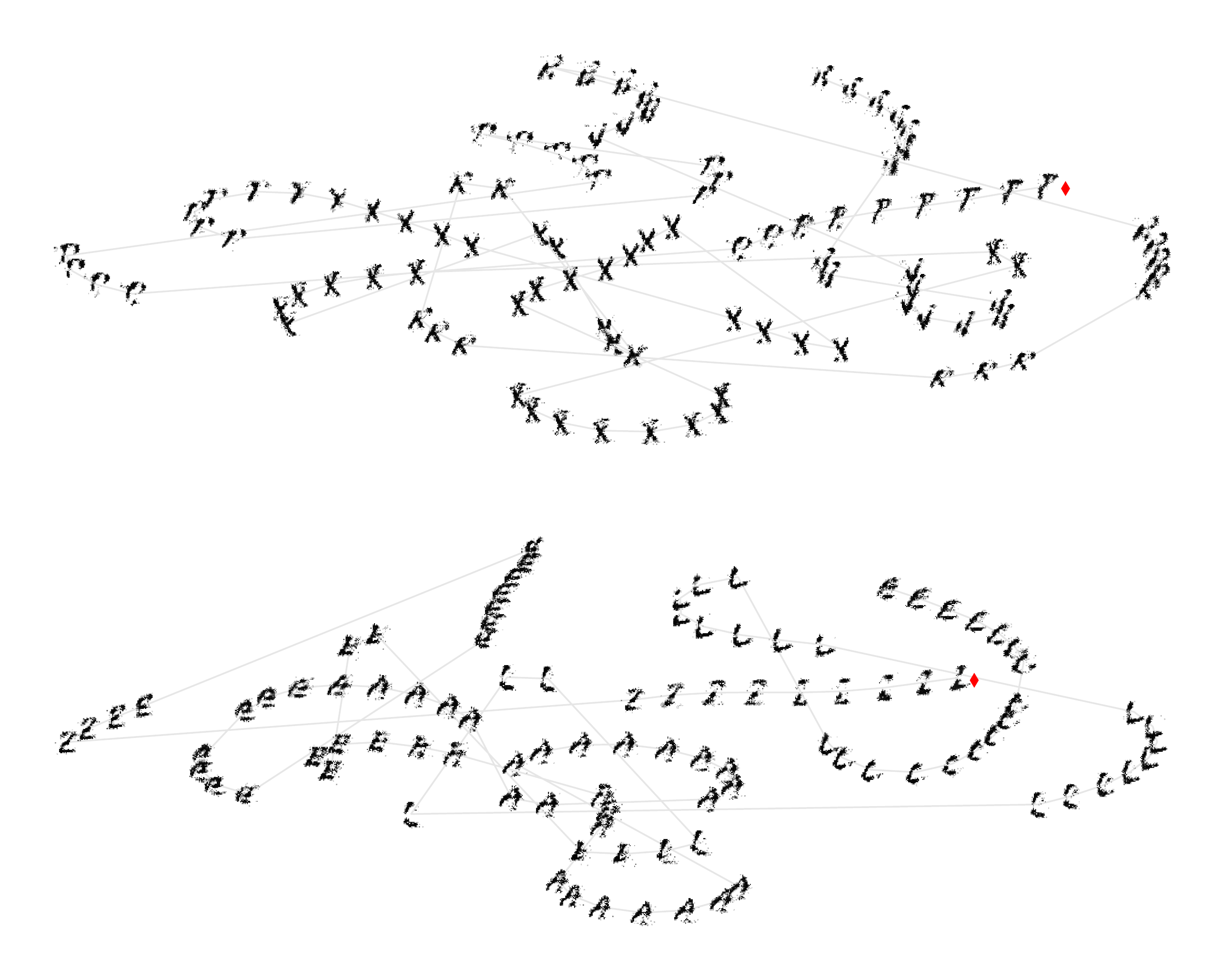}
    \caption{
        t-SNE representation \cite{maaten2008visualizing} of consecutively generated images of two of the four SSNs trained on EMNIST. 
        Both SSNs smoothly traverse several regions of the state space representing image classes while dreaming.
       	The red diamond marks the first image in the sequence, gray lines connect consecutive images.
       	Consecutive images are \SI{200}{ms} apart.
}\label{fig:figS4}
\end{figure*}
\vspace*{\fill}
\clearpage
\vspace*{\fill}
\begin{figure*}[h]
    \centering
    \includegraphics[width=\textwidth]{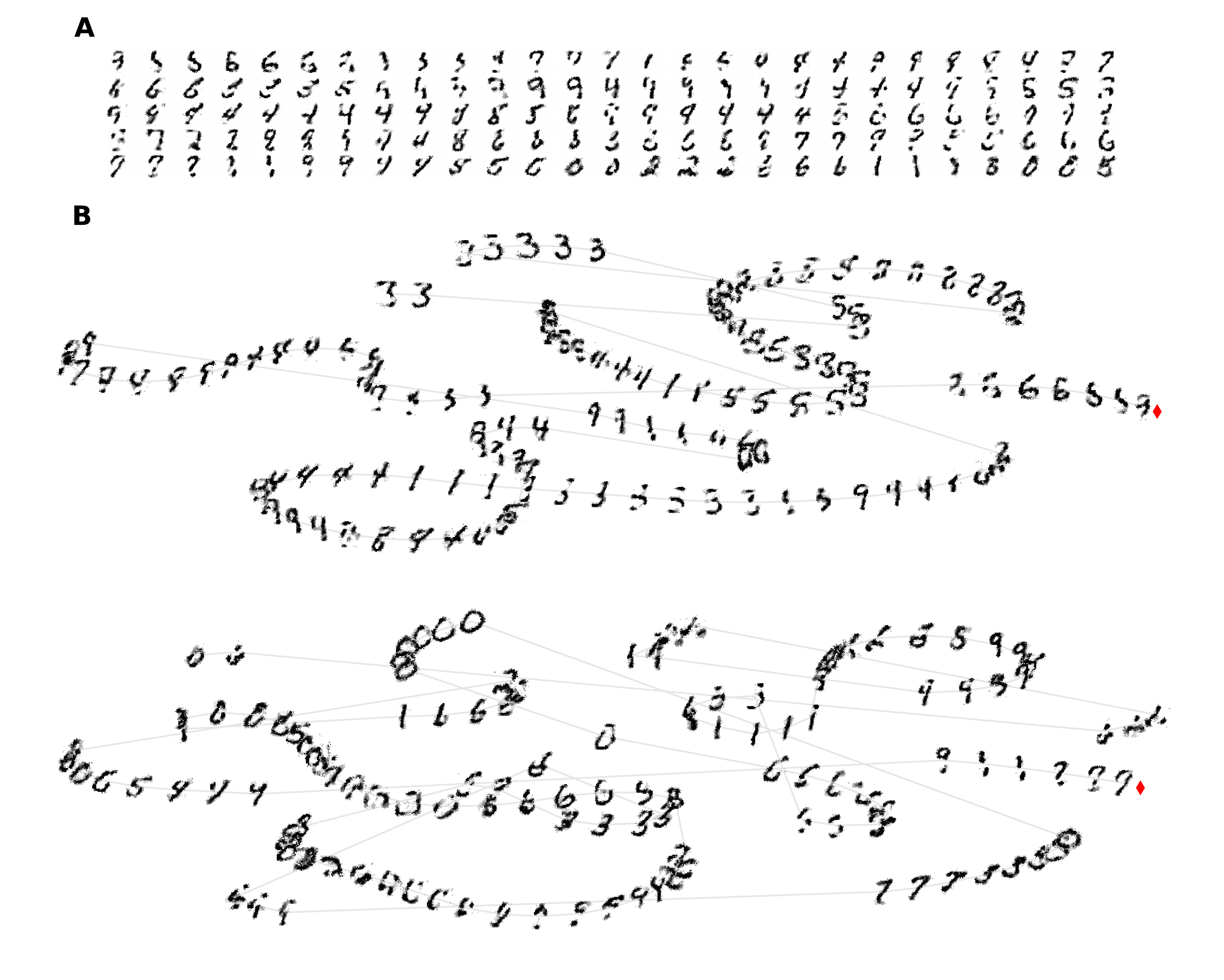}
    \caption{
        Dreaming of MNIST in an ensemble of networks. \textbf{(A)} Dreaming ensemble of five hierarchical SSNs with 784 visible, 500 hidden and 10 label neurons (without explicit noise).
        Each row represents samples from a single network of the ensembles, with samples being \SI{375}{ms} apart.
        To set up the ensemble, a restricted Boltzmann machine was trained on the MNIST dataset and the resulting parameters translated to corresponding neurosynaptic parameters of the ensemble.
        Here, to facilitate mixing, we used short-term depression to modulate synaptic interactions and weaken attractor states that would be otherwise difficult to escape \cite{leng2018spiking}.
        \textbf{(B)} t-SNE representation \cite{maaten2008visualizing} of consecutively generated images of two of the five SSNs trained on MNIST digits.
      	Both SSNs are able to generate and mix between diverse images of different digit classes while dreaming.
      	The red diamond marks the first image in the sequence, gray lines connect consecutive images.
      	Consecutive images are \SI{400}{ms} apart.
        }\label{fig:figS5}
\end{figure*}
\vspace*{\fill}
\clearpage
\vspace*{\fill}
\begin{figure*}[h]
    \centering
    \includegraphics[width=\textwidth]{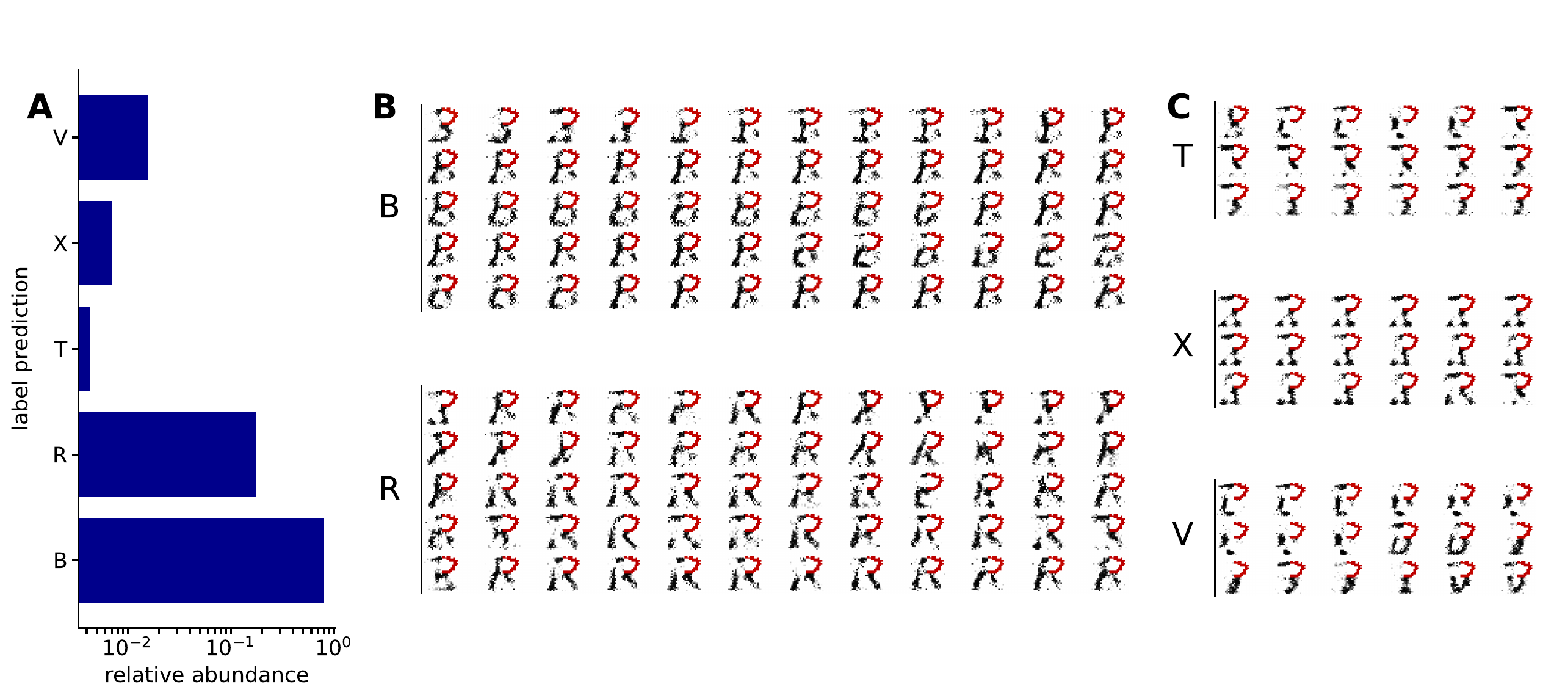}
    \caption{Pattern completion in an ensemble trained on EMNIST. \textbf{(A)} Relative abundance of the label output while clamping parts of a "B". 
    Most of the time (\textbf{79.85\%}), the image is correctly classified as a "B". 
    The closest alternative explanation, an "R", is generated second most (\textbf{17.45\%}). 
    The remaining classes are explored significantly less often by the network (\textbf{0.43\%, 0.70\%, 1.57\%}).
    \textbf{(B)} Examples of the visible layer activity while the label layer classifies the partially clamped images either as a "B" (top) or an "R" (bottom).
    \textbf{(C)} Examples of the visible layer activity while classifying the image as a "T", "X" or "V".
    In these cases, the images generated by the visible neurons show prominent features of these letters.}\label{fig:figS6}
\end{figure*}
\vspace*{\fill}
\clearpage
\vspace*{\fill}
\begin{figure*}[h]
    \centering
    \includegraphics[width=\textwidth]{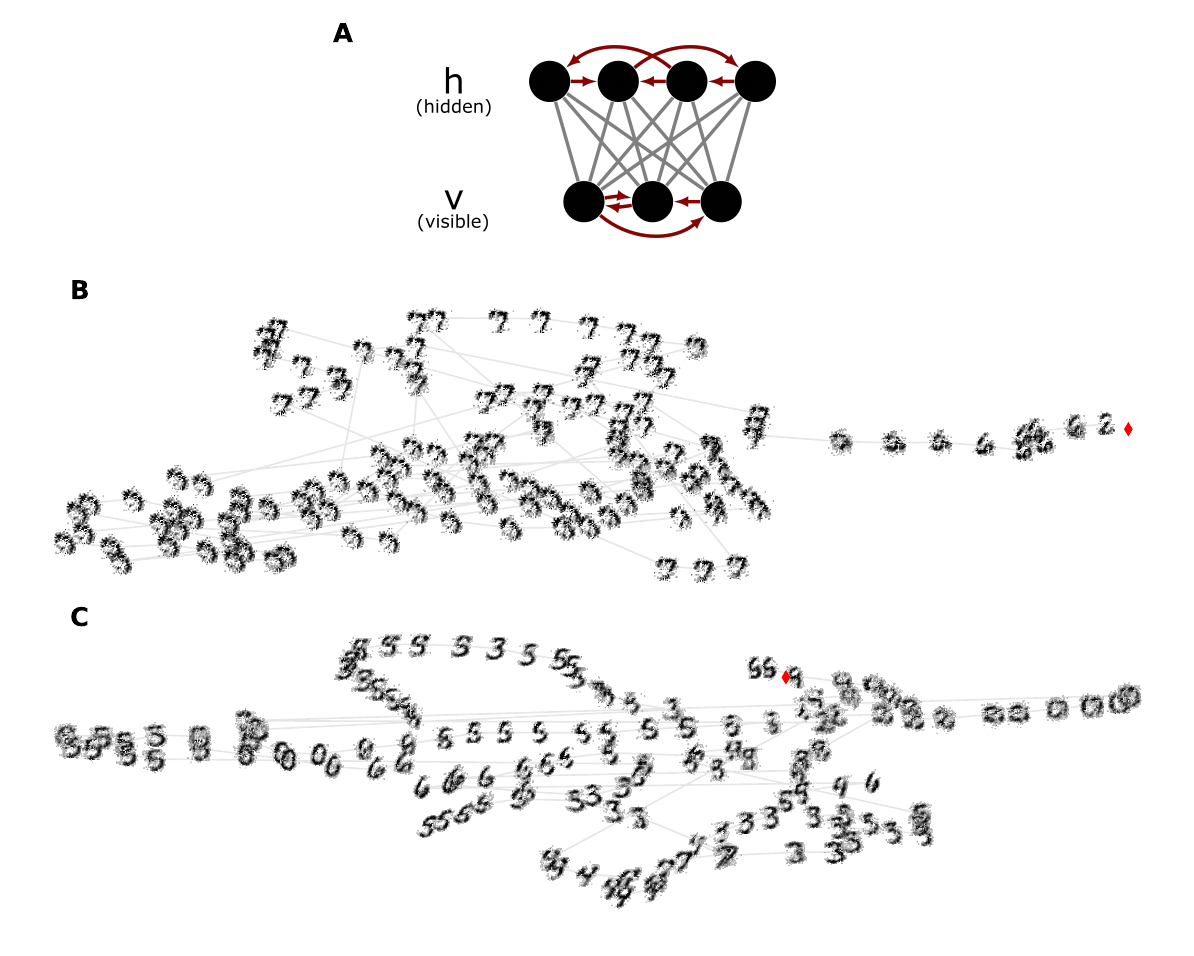}
    \caption{A single hierarchical network with 784-200 (visible-hidden) neurons generating samples of the MNIST handwritten digits dataset without explicit noise sources (in simulation), represented via t-SNE.
    \textbf{(A)} Illustration of the used network architecture. Lateral (non-plastic) connections in each layer were utilized as a noise source (red), with an interconnectivity of $\epsilon = 0.2$. 
    \textbf{(B,C)} Averaged activity (average window $\pm 90$ms) of the visible layer (B) after initializing the network and (C) after further training the network. After initialization, the network is able to generate recognizable images but does not mix well between different digit classes since the network is not able to correctly utilize its own background activity as noise yet (B). Further training the network on images of the MNIST training set (using standard Contrastive Divergence with batch size $100$, learning rate $\frac{40}{t+2000}$ with $t$ the number of updates, for $1000$ training updates and a presentation time per training sample of $200$ms) improves both image quality and mixing (C). During the second training phase, neurosynaptic parameters are adjusted such that every neuron is able to perform its task with the available background activity it receives.}\label{fig:figS7}
\end{figure*}
\vspace*{\fill}